\documentclass[]{interact}
\usepackage{lscape}
\usepackage{epstopdf}
\usepackage{subfigure}
\usepackage[numbers,sort&compress]{natbib}
\bibpunct[, ]{[}{]}{,}{n}{,}{,}
\renewcommand\bibfont{\fontsize{10}{12}\selectfont}
\makeatletter
\newcommand\extralabel[2]{{\edef\@currentlabel{\@currentlabel#2}\label{#1}}}
\makeatother
\usepackage{graphicx}
\usepackage{array}
\usepackage{bm}
\usepackage{float}
\theoremstyle{plain}
\newtheorem{theorem}{Theorem}[section]

\newtheorem{corollary}[theorem]{Corollary}
\newtheorem{proposition}[theorem]{Proposition}
\newtheorem{Remark}{Remark}
\theoremstyle{definition}

\theoremstyle{remark}

\makeatletter
\newcommand*{\rom}[1]{\expandafter\@slowromancap\romannumeral #1@}
\makeatother
\usepackage{hyperref}
\hypersetup{
   colorlinks=true,
   linkcolor=blue,
   filecolor=magenta,      
   urlcolor=cyan,
   pdftitle={Overleaf Example},
   pdfpagemode=FullScreen,
   }

\def\tr{\mathop{\rm tr}\nolimits}

\begin{document}

\title{A Computational Note on the Graphical Ridge in High-dimension}

\author{
\name{A. Bekker\textsuperscript{a,b}\thanks{CONTACT A. Bekker. Email: andriette.bekker@up.ac.za}, A. Kheyri\textsuperscript{a} and M. Arashi\textsuperscript{c}}
\affil{\textsuperscript{a}Department of Statistics, Faculty of Natural and Agricultural Sciences, University of Pretoria, Pretoria, South Africa\\
\textsuperscript{b}Centre of Excellence in Mathematical and Statistical Sciences, Johannesburg, South Africa
\textsuperscript{c}Department of Statistics, Faculty of Mathematical Sciences, Ferdowsi
University of Mashhad, Mashhad, Iran}
}


\maketitle

\begin{abstract}
This article explores the estimation of precision matrices in high-dimensional Gaussian graphical models. We address the challenge of improving the accuracy of maximum likelihood-based precision estimation through penalization. Specifically, we consider an elastic net penalty, which incorporates both $L_1$ and Frobenius norm penalties while accounting for the target matrix during estimation.
To enhance precision matrix estimation, we propose a novel two-step estimator that combines the strengths of ridge and graphical lasso estimators. Through this approach, we aim to improve overall estimation performance. Our empirical analysis demonstrates the superior efficiency of our proposed method compared to alternative approaches.
We validate the effectiveness of our proposal through numerical experiments and application on three real datasets. These examples illustrate the practical applicability and usefulness of our proposed estimator. 
\end{abstract}


\section{Introduction}
In recent years, there has been a substantial amount of research dedicated to investigating the estimation of the covariance matrix and its inverse, commonly known as the precision matrix.
The motivation behind these studies arises from the critique that commonly employed likelihood-based approaches yield inaccurate estimates for the covariance and precision matrices, particularly in scenarios where the number of variables exceeds the available data points.
One commonly employed approach is incorporating an $L_1$ penalty into the log-likelihood function.
Nearly concurrently, \cite{yuan2007model}, \cite{banerjee2008model}, \cite{friedman2008sparse} all investigated the estimation of the precision matrix, represented by a graphical lasso. This method has attracted a lot of interest and turned into an active topic of study due to the sparse solution (\cite{fan2009network}; \cite{bien2011sparse}; \cite{witten2011new}; \cite{mazumder2012graphical}; \cite{danaher2014joint}; \cite{avagyan2017improving}). \\
While sparsity in precision matrix estimation is often desirable, there are situations where precise representations of the high-dimensional precision matrix are fundamentally advantageous. Moreover, it is important to note that the true precision matrix does not necessarily need to exhibit sparsity, meaning it may not have a significant number of zero elements.
In such scenarios, a regularization approach is employed to effectively shrink the estimated elements of the precision matrix. Several studies have explored the use of ridge regularization in the log-likelihood function by incorporating a Frobenius norm penalty. Relevant works in this area include \cite{ledoit2004well}; \cite{schafer2005shrinkage}; \cite{warton2008penalized}; \cite{van2016ridge}; \cite{van2019generalized}, among others.\\
A challenge in the regularized precision estimation is to find a closed-form solution dealing with the $L_1$ penalty. Indeed element-wise estimator is not the case, unlike vector-variate estimation. This paper considers a combination of alternative ridge and $L_1$ penalty in a 2-step approach. The main goal of this paper is to present and compare the efficiency gained by considering our proposed methodology for estimating the precision matrix in the Gaussian graphical model compared to some existing methods. Secondly, we propose a closed-form
estimator for the precision matrix when we have an elastic net penalty. Thus, we organize the rest of this article as follows. In Section \ref{se2}, we review two types of the archetypal ridge and the alternative ridge estimators. Then, we propose our 2-step approach with elastic net type penalty to find a non-sparse solution for the precision matrix in \ref{se3}. In the numerical study, Section \ref{se4} confirms that the proposed estimation is comparable with some alternative estimations, and in most cases, it behaves better. Section \ref{se5} provides examples of real-world data applications strengthening our methodology and argument. To provide a quick summary, we discuss our key findings briefly in Section \ref{sec6}. 
\section{Existing Methods}\label{se2}
Gaussian graphical models are widely popular in graphical modeling due to their ease of handling. One key advantage of Gaussian graphical models is that both their conditional and unconditional dependence structures are fully determined by the precision matrix of the underlying distribution. Estimating the covariance matrix in such models can be achieved using standard methodologies, such as maximum likelihood (ML) estimation or its regularized variants.\\
Mathematically, a Gaussian graphical model is represented by a pair $(\mathbf{z}, \mathcal{G})$, where $\mathbf{z}$ is a random vector following a multivariate normal distribution $\mathcal{N}_p(\bm\mu, \bm\Sigma)$. Here, $\bm\mu \in \mathbb{R}^p$ represents an arbitrary mean vector, and $\bm\Sigma$ is a symmetric and positive definite covariance matrix denoted as $\bm\Sigma \succ 0$. The graph $\mathcal{G}$ corresponds to the structural representation of the conditional independence relationships among the variables.\\
In Gaussian graphical models, the graph $\mathcal{G}$ and the precision matrix $\bm\Theta:=\bm\Sigma^{-1}$ are closely interconnected. The zero entries in the precision matrix, denoted by $\theta_{ij}=0$, correspond to pairs of variables $\mathbf{z}_i$ and $\mathbf{z}_j$ that are conditionally independent given all the other variables, i.e., $\mathbf{z}_i\mathbf{z}_j|\mathbf{z}_{\{1,\ldots,p\}\setminus i,j}$. In other words, there is no edge between these two variables (nodes) in the graph $\mathcal{G}$.
Additionally, two variables $\mathbf{z}_i$ and $\mathbf{z}_j$ are independent if and only if there is no path of edges connecting them in the graph $\mathcal{G}$.

The maximum likelihood (ML) estimation is an analytical technique used to estimate the precision matrix $\bm\Theta$. The ML estimator for $\bm\Theta$ is typically expressed as
\begin{equation}\label{eq1}
\hat{\bm\Theta}=\arg\underset{\bm\Theta\succ 0}{\max}  \{\log\det (\bm\Theta)-\tr(\bm S\bm\Theta)\},
\end{equation}
Where $\bm S=\mathbf{z}^T\mathbf{z}/n$ represents the sample covariance matrix, the ML estimator of $\bm \Theta$ is given by $\bm S^{-1}$. However, when applying this technique to estimate $\bm \Theta$, two issues may arise. Firstly, in the high-dimensional case where the number of variables $p$ exceeds the number of observations $n$, the empirical covariance matrix $\bm S$ becomes singular, preventing the inversion required to estimate $\bm \Theta$. Even if $p$ and $n$ are approximately equal, and $\bm S$ is non-singular, the maximum likelihood estimate for $\bm \Theta$ tends to have a relatively high variance. Secondly, in a graphical model, it is often desirable to identify pairs of variables that are disconnected and conditionally independent, which corresponds to zero entries in the precision matrix $\bm \Theta$. However, the maximum likelihood estimation given by equation \eqref{eq1} generally does not yield any zero elements in the estimated $\bm \Theta$.
\subsection{Ridge Estimators}
To address the limitations mentioned above, the likelihood function is augmented with a penalty function in high-dimensional settings, leading to the regularized maximum likelihood estimator. In the case of non-sparse solutions, a ridge estimator based on archetypes was proposed in \cite{ledoit2004well}. The corresponding penalized log-likelihood can be expressed as follows:
\begin{equation}\label{eq2}
\log\det (\bm \Theta)-(1-\lambda)\tr(\bm S\bm \Theta)-\lambda \tr(\bm \Theta\bm \Gamma),
\end{equation}
where $\lambda\in\text{(0,1]}$ is the tuning parameter, and $\bm \Gamma$ is a positive definite target matrix. Refer to \cite{arashi2019theory} for more information about the ridge notion. A common target selection is a diagonal matrix, where its elements are the same as the corresponding elements of the sample covariance matrix $\bm{S}$ in \eqref{eq2}.\\
Solving the optimization problem \eqref{eq2}, we have the first archetypal ridge estimator given by
\begin{equation}\label{eq002}
\hat{\bm \Theta}^1(\lambda)=[(1-\lambda)\bm S+\lambda\bm \Gamma]^{-1}.   
\end{equation}
In the second archetypal ridge approach, Schafer and Strimmer \cite{schafer2005shrinkage} introduced a specific form of penalized log-likelihood. They considered the following expression:
\begin{equation}\label{eq3}
\log\det (\bm \Theta)-\tr(\bm S\bm \Theta)-\lambda \tr(\bm \Theta),
\end{equation}
where $\lambda\in (0,\infty)$ and the corresponding estimator has form
\begin{equation*}
\hat{\bm \Theta}^2(\lambda)=[\bm S+\lambda I_p]^{-1}.    
\end{equation*}
Contrary to our expectations, the penalties in equations \eqref{eq2} and \eqref{eq3} do not correspond to the Frobenius norm penalty, which is defined as the square root of the sum of squares of the elements of the precision matrix.\\
The alternative ridge approach introduced by \cite{van2016ridge} and the ridge-type operator for precision matrix estimation (ROPE) proposed by \cite{kuismin2017precision} both utilize the Frobenius norm penalty. These methods consider two different types of ridge precision estimators, where the penalized log-likelihood function is given by
\begin{equation}\label{eq4}
\log\det (\bm \Theta)-\tr(\bm S\bm \Theta)-\frac{1}{2}\lambda ||\bm \Theta-\bm T||_F^2\},
\end{equation}
with $\bm T$ denoting a symmetric positive definite target matrix and the  penalty parameter $\lambda\in (0,\infty)$. The type-I ridge precision estimator (or ROPE), the consequence of the optimization problem \eqref{eq4}, is given by
\begin{equation}\label{eq5}
\hat{\bm \Theta}^{\rom{1}a}(\lambda)=\left[\left(\lambda \bm I_p+\frac{1}{4}(\bm S-\lambda  \bm T)^2\right)^{1/2}+\frac{1}{2}(\bm S-\lambda \bm T)\right]^{-1}.
\end{equation}
The RHS and LHS limits of $\hat{\bm \Theta}^{\rom{1}a}$ are the inverses of the ML estimator $\bm S$ (if $p<n$) and the target matrix, respectively. Target matrix $\bm T$ is considered positive definite to avoid having wrong estimates in the limit.
\begin{Remark}
The type-I ridge estimator can also be represented without inversion as
\begin{equation}\label{eq6}
\hat{\bm \Theta}^{Ia}(\lambda)=\frac{1}{\lambda}\left[\left(\lambda \bm I_p+\frac{1}{4}(\bm S-\lambda  \bm T)^2\right)^{1/2}-\frac{1}{2}(\bm S-\lambda \bm T)\right].
\end{equation}
\end{Remark}
Also, the type-II ridge estimator is obtained by considering $\bm T=\bm 0$ in \eqref{eq5} (or \eqref{eq6}) as follows
\begin{equation}\label{eq7}
\hat{\bm \Theta}^{\rom{2}a}(\lambda)=\left[\left(\lambda \bm I_p+\frac{1}{4}\bm S^2\right)^{1/2}+\frac{1}{2}\bm S\right]^{-1}.
\end{equation}
The RHS and LHS limits of $\hat{\bm \Theta}^{\rom{2}a}$, like the second archetypal estimator, are the inverse of the ML estimator $\bm S$ (where $p<n$) and the null matrix, respectively. It should be mentioned that \cite{witten2009covariance} studied the estimator \eqref{eq7} in a different context. \\
In a recent study by \cite{van2019generalized}, a generalized ridge precision estimator was proposed, allowing for entry-wise penalization. This approach enables the estimator to shrink towards a user-specified, nonrandom target matrix while maintaining positive definiteness and consistency. Furthermore, the authors derived a generalization of the graphical lasso estimator and its elastic net counterpart. It is worth noting that Kovács et al., \cite{kovacs2021graphical} considered elastic net type penalization for precision matrix estimation, specifically in the presence of a diagonal target matrix. They demonstrated that adapting the iterative procedure proposed by \cite{van2019generalized} for the elastic net problem is feasible but may not be computationally efficient.\\
All of the above set the platform for the motivation of this paper. Here, in this paper,
we consider the elastic net type penalty and propose a novel non-sparse estimate technique for
precision matrix using a 2-step strategy.

\section{Constructing the Proposed Estimator}\label{se3}
In this section, we begin by providing the motivation for our methodology. Considering the continuity property of ridge estimators and disregarding the influence of sparsity, there is a possibility that an estimator of $\bm \Theta$ may deviate significantly from the true value $\Theta$; for instance, the estimators mentioned in Section 2. Hence, it is of major importance to find a closer estimator. Therefore, we can penalize the deviance between any primary or initial target estimator and $\bm \Theta$ and tune it to improve the final estimation using the Frobenius norm penalty. Then we add the $L_1$ penalty, returning to the well-known elastic net regularization framework, but with refinement. 

Given a sample $\bm X_{n\times p}$ of $n$ realizations from a $p$-variate Gaussian distribution with zero mean and positive definite covariance matrix $\bm \Sigma$. Consider the following optimization problem aimed at estimating the unknown precision matrix based on $n$ observations:
\begin{equation}\label{eq8}
    \hat{\bm\Theta}(\alpha,\lambda,\bm T)=\arg\underset{\bm\Theta\succ 0}{\max} \{\log\det (\bm\Theta)-\tr(\bm S\bm\Theta)-\lambda(\alpha||\bm\Theta||_1+\frac{(1-\alpha)}{2}||\bm\Theta-\bm T||_F^2)\},
    \end{equation}
where $\bm T$ is a known positive semi-definite diagonal target matrix. Also $\lambda\geq 0 $ and $\alpha\in [0,1]$ are tuning parameters. The target $\bm T$ can be any of the estimators mentioned in Section \ref{se2} or a sparse version like the graphical lasso. 

We can rewrite the optimization problem \eqref{eq8} as
\begin{equation}\label{e01}
    \underset{\bm\Theta\succ 0}{max}\underset{\parallel U\parallel 	_\infty\leq \lambda\alpha}{min}\log\det (\bm\Theta)-\tr(\bm\Theta,\bm S+\bm U)-\frac{\lambda(1-\alpha)}{2}||\bm\Theta-\bm T||_2^2,
\end{equation}
where we express the $L_1$ norm as
\begin{equation}
   ||\bm\Theta ||_1= \underset{\parallel \bm U\parallel 	_\infty\leq 1}{max}tr(\bm\Theta \bm U),
\end{equation}
and $||\bm U||_\infty$ represents the maximum absolute value element of the symmetric matrix $\bm U$.
To achieve a closed-form solution for \eqref{e01}, we have the following results. All the proofs are
provided in Appendix A.
\begin{proposition}\label{prop1}
The optimization problem \eqref{e01} can be reformulated as an equivalent expression, which is given by
\begin{eqnarray}\label{e00002}
\nonumber\underset{\parallel \bm U\parallel 	_\infty\leq \lambda\alpha}{min}\log\det \left(\frac{1}{\lambda(1-\alpha)}\left[\left(\mathcal{\bm A}^2+\lambda(1-\alpha)\bm I_p\right)^{1/2}-\mathcal{\bm A}\right]\right)\\
-\tr\left(\frac{1}{\lambda(1-\alpha)}\left[\left(\mathcal{\bm A}^2+\lambda(1-\alpha)\bm I_p\right)^{1/2}-\mathcal{\bm A}\right]\mathcal{\bm A}\right).
\end{eqnarray}
where $\mathcal{\bm A}=\frac{1}{2}(\bm S+\bm U-\lambda(1-\alpha)\bm T)$.
\end{proposition}
\begin{corollary}
Since for any square nonsingular matrix $\bm A$ we have $\log\det (\bm A)=\tr(\log (\bm A))$ so the optimization problem \eqref{e00002} can also represented as follow
\begin{eqnarray}
\underset{\parallel \bm U\parallel 	_\infty\leq \lambda\alpha}{min}\tr(\log (\mathcal{\bm B})-\mathcal{\bm B}\mathcal{\bm A}),
\end{eqnarray}
where $\mathcal{\bm B}=\frac{1}{\lambda(1-\alpha)}[\left(\mathcal{\bm A}^2+\lambda(1-\alpha)\bm I_p\right)^{1/2}-\mathcal{\bm A}]$.
\end{corollary}
Also, we can express proposition \ref{prop1} concerning eigenvalues of matrix $\bm S+\bm U$. We summarize the result in the following proposition.
\begin{proposition}\label{prop2}
Let $\bm S+\bm U=\bm P\bm D\bm P^{-1}$ where $\bm D$ is a diagonal matrix containing the eigenvalues of $\bm S+\bm U$ on the diagonal, and $\bm P$ is a matrix that consists of the corresponding eigenvectors as its columns. Also consider $\bm T=\gamma \bm I_p$ with $\gamma\in (0,\infty)$, Substituting this decomposition, we can rewrite the optimization problem \eqref{e00002} as follows
\begin{eqnarray*}
\underset{\parallel \bm U\parallel_\infty\leq \lambda\alpha}{max} \sum_{i=1}^p\log \left(\sqrt{(b_i)^2+\lambda(1-\alpha)}+b_i\right)+\frac{1}{\lambda(1-\alpha)}b_i\left(\sqrt{(b_i)^2+\lambda(1-\alpha)}-b_i\right),
\end{eqnarray*}
where $b_i=1/2(\lambda_i-\lambda(1-\alpha)\gamma$ and $\lambda_1,\cdots,\lambda_p$ are eigenvalues of the matrix $\bm S+\bm U$.
\end{proposition}
\subsection{Conceptualizing the Estimator through Conjecture}
Our problem statement starts from where we want to find an analytical solution for Proposition \ref{prop1}. Indeed, the solution can be obtained numerically. Here, we present an intuitive way to obtain a closed-form solution.\\ Noticing the proof of Proposition \ref{prop1}, in step one, we solve the objective function subject to $||\bm U||_\infty\leq \lambda\alpha$ by setting the tuning parameter equal to $\alpha\lambda$ resulting in the glasso. Then, in step two, we substitute the glasso into the objective function and solve for $\bm \Theta$. Subsequently with the following closed-form solution is obtained 
\begin{equation}\label{eq13}
\hat{\bm \Theta}^{2s}(\alpha,\lambda)=\left[\left(\lambda \bm I_p+\frac{1}{4}(\bm W-\lambda(1-\alpha) \bm T)^2\right)^{1/2}+\frac{1}{2}(\bm W-\lambda(1-\alpha) \bm T)\right]^{-1},
\end{equation}
where $\bm W$ is the covariance matrix estimate obtained from the glasso with the tuning parameter $\alpha\lambda$. 
In Appendix, we have conducted a toy simulation study to show how much the numerical solution is close to the 2-step ridge estimator $\hat{\bm \Theta}^{2s}$ in \eqref{eq13}. 

\subsection{Detailed Development of the Estimator}
The estimator presented in \eqref{eq13} can be achieved by considering the following penalized log-likelihood:
\begin{equation}\label{eq15}
\log\det (\bm \Theta)-(1-\lambda_1)\tr(\bm S\bm \Theta)-\lambda_1 \tr(\bm \Theta\bm \Gamma)-\frac{1}{2}\lambda_2 ||\bm \Theta-\bm T||_F^2,
\end{equation}
where $\lambda_1\in (0,1]$ and $\lambda_2$ is a non-negative tuning parameter. The penalized log-likelihood \eqref{eq15} is a combination of first archetype ridge penalized log-likelihood, \eqref{eq2} and alternative ridge in \eqref{eq4}. Setting $\lambda_1=0$ yields the alternative ridge estimator, whereas $\lambda_2=0$ esults in the Ledoit-Wolf estimator, as detailed in Equation \eqref{eq002}. 
To address the corresponding convex optimization problem, we consider the normal equations derived from the derivative of Equation \eqref{eq15} with respect to $\bm \Theta$,
\begin{equation}
    \bm \Theta^{-1}-(1-\lambda_1)\bm S-\lambda_1 \bm\Gamma -\lambda_2 (\bm\Theta-\bm T).
\end{equation}
Following the method outlined in Kuismin (2017), we arrive at the following solution for the estimator:
\begin{equation}\label{eq18}
    \hat{\bm\Theta}=\left[\left(\lambda_2 \bm I_p+\frac{1}{4}((1-\lambda_1)\bm S+\lambda_1 \bm\Gamma-\lambda_2 \bm T)^2\right)^{1/2}+\frac{1}{2}((1-\lambda_1)\bm S+\lambda_1 \bm\Gamma-\lambda_2\bm T)\right]^{-1}.
\end{equation}
This result indicates an estimator similar to the Alternative Ridge, but with a key difference: $\bm S$ in the equation is replaced with a convex combination of $\bm S$ and $\bm \Gamma$. Notably, setting $\lambda_1=1$ and using the target matrix $\Gamma$ via the graphical lasso approach, we achieve the estimator outlined in Equation \eqref{eq13}. The estimator mentioned in Equation \eqref{eq18}, to the best of my knowledge, has not yet been explored in existing literature. This opens a potential avenue for future research in this domain. However, the primary focus of this paper remains on our proposed estimator. In the subsequent section, we will illustrate the effective performance of this estimator, underscoring its practical utility and significance in the field.

\section{Simulation Study}\label{se4}
In this section, we evaluate and compare the performance of our proposed 2-step estimator $\hat{\bm \Theta}^{2s}$ given in \eqref{eq13} with other existing methods, namely, graphical lasso or \textbf{Glasso} (\cite{friedman2008sparse}), type-I and II ridge or \textbf{Alternative ridge} estimators (\cite{kuismin2017precision} and \cite{van2016ridge}), and the graphical elastic net or \textbf{Gelnet} precision matrix estimator (\cite{kovacs2021graphical}). The simulation structure used in our evaluation is based on the framework proposed by \cite{kuismin2017precision}.

We we generate simulated data from a multivariate Normal distribution $\mathcal{N}_p(\mathbf{0},\bm \Sigma)$, where $\bm \Sigma=[\sigma_{k,k^{'}}]$ and $\bm \Theta=[\theta_{k,k^{'}}]$ are positive definite matrices of size $p\times p$. In our comparison, we consider six different models representing network structures. The methods are tested on these structures, considering both sparse and non-sparse scenarios. We specifically consider network structures with $p=20, 50, 100$ nodes, and a sample size of $n=50$. 
\begin{itemize}
    \item Network 1. One of the models used in the comparison is a compound symmetry model, where $\sigma_{k,k}=1$ for all $k$, indicating equal variances across variables, and $\sigma_{k,k'}=0.6^2$ for $k\neq k'$, indicating a constant covariance between any pair of variables. This covariance matrix exhibits a structured and non-sparse pattern.
    \item Network 2. The precision matrix $\bm \Theta$ is derived from the prototype matrix $\bm \Theta_0$ by standardizing it to have a unit diagonal. The prototype matrix $\bm \Theta_0$ is defined as $\bm \Theta_0=\bm A+a\bm I_p$, where each off-diagonal entry in $\bm A$ is generated independently. Specifically, each off-diagonal entry is set to $0.5$ with a probability of $0.1$, indicating a non-zero relationship, or set to $0$ with a probability of $0.9$, indicating no relationship. The parameter $a$ is chosen such that the condition number of the matrix is equal to $p$, ensuring a well-conditioned matrix. This precision matrix exhibits both unstructured and sparse patterns
    \item Network 3. The precision matrix $\bm \Theta$ is estimated as $\bm \Theta=\frac{1}{n}\bm Y^T\bm Y$, where $\bm Y=[y_{i,j}]$ is an $n\times p$ matrix with $n=10,000$ observations and $p$ variables. Each element $y_{i,j}$ is drawn independently from a standard normal distribution $\mathcal{N}(0,1)$. The resulting precision matrix $\bm \Theta$ is an estimate of the true precision matrix and represents the pairwise dependencies among the variables. In this case, the precision matrix is considered unstructured and non-sparse, as there are no specific patterns or constraints imposed on its elements.
    \item Network 4. The precision matrix $\bm \Theta$ follows a star model, where $\theta_{k,k}=1$ for all variables. Additionally, there is a specific pattern of connections between the first variable and all other variables, with $\theta_{1,k}=\theta_{k,1}=0.1$, indicating a connection between the first variable and the rest. All other off-diagonal elements $\theta_{k,k'}$ are set to zero, implying that there are no connections between variables other than the first variable. This results in a structured and sparse precision matrix, as it exhibits a specific pattern of nonzero elements along with many zero elements.
    \item Network 5. The covariance matrix $\bm \Sigma$ follows a moving average (MA) model, where each diagonal element $\sigma_{k,k}$ is equal to 1, indicating that each variable has a unit variance. The off-diagonal elements follow a specific pattern: $\sigma_{k,k-1}=\sigma_{k-1,k}=0.2$, representing a weaker correlation between adjacent variables, and $\sigma_{k,k-2}=\sigma_{k-2,k}=0.2^2$, representing an even weaker correlation between variables that are two steps apart. All other off-diagonal elements are set to zero, indicating no correlation beyond the specified patterns. This results in a structured and sparse covariance matrix, as it exhibits a specific pattern of nonzero elements along with many zero elements.
    \item Network 6. The covariance matrix $\bm \Sigma$ follows a diagonally dominant model. It is constructed based on a matrix $\bm B$ derived from another matrix $\bm A$ by $\bm B=\frac{1}{2}(\bm A+\bm A^T)$, where $\bm A$ is a $p\times p$ matrix. Each off-diagonal element $a_{k,k^{'}}$ (for $k \neq k^{'}$) in $\bm A$ is randomly drawn from a uniform distribution $U(0,1)$, and the diagonal elements $a_{k,k}$ are set to zero. To obtain the matrix $\bm D$, which is used to scale $\bm B$, each element of $\bm B$ is divided by a scaling factor $\gamma$. The scaling factor $\gamma$ is determined as the largest row sum of the absolute values of the elements in $\bm B$. Thus, $\bm D$ is computed as $\bm D = \frac{1}{\gamma}\bm B$, resulting in a matrix $\bm D$ with values between 0 and 1. The final step involves constructing the covariance matrix $\bm \Sigma$. The off-diagonal elements $\sigma_{k,k^{'}}$ are set to the corresponding elements $d_{k,k^{'}}$ from $\bm D$. The diagonal elements $\sigma_{k,k}$ are computed as $1 + e_i$, where $e_i$ is randomly drawn from a uniform distribution $U(0,0.1)$. This ensures that the diagonal elements have a slightly larger value than 1, introducing some variability. Overall, this construction results in an unstructured and non-sparse covariance matrix, where the off-diagonal elements follow a specific pattern based on the values in $\bm D$ and the diagonal elements have a slight perturbation.
\end{itemize}
To assess the performance of all methods, we conduct 100 independent simulations for each network structure and calculate the average of various loss functions. We determine the optimal tuning parameter $\lambda$ for each method by performing five-fold cross-validation on each simulation run. For the Glasso, Gelnet, and Alternative ridge methods, we utilize the five-fold cross-validation provided by the R-package "GLassoElnetFast" with $\lambda$ values ranging from $0$ to $10$ and $\alpha$ values ranging from $0$ to $1$. As target matrices, we consider both the identity matrix $\bm I_p$ and a scalar matrix $\nu \bm I_p$, where $\nu$ is calculated as $\nu = \frac{p^2}{\text{tr}(\bm S)}$.
\subsection{Performance Measures}
For the evaluation of the performance of a given estimator $\hat{\bm \Theta}$, we consider three loss functions which have been widely used (see, e.g., \cite{avagyan2017improving}, \cite{kuismin2017precision}, \cite{bernardini2021new} and \cite{kovacs2021graphical}).
Here are the four loss functions:
\begin{enumerate}
    \item Kullback-Leibler loss (KL): This loss function measures based on the likelihood function. It is defined as $KL = \text{tr}(\bm \Sigma \hat{\bm \Theta}) - \text{logdet}(\bm \Sigma \hat{\bm \Theta}) - p$.
    \item  L2 loss (L2): The L2 loss calculates the Frobenius norm of the difference between the estimated precision matrix and the true precision matrix. It is given by $L2 = ||\bm \Theta - \hat{\bm \Theta}||_F$.
    \item Quadratic loss (QL): It is defined as $QL = \text{tr}((\bm \Sigma \hat{\bm \Theta} - \bm I_p)^2)$, where $\bm I_p$ is the identity matrix of size $p$.
    \item Spectral norm loss (SP): The spectral norm loss measures the largest singular value of the matrix $(\bm \Theta - \hat{\bm \Theta})^2$. 
\end{enumerate}
 These loss functions provide quantitative measures to evaluate the performance of the estimator by capturing different aspects of the dissimilarity between the estimated precision matrix and the true precision matrix. 
 
\begin{figure}[h!]
    \centering
    \includegraphics[scale=0.42]{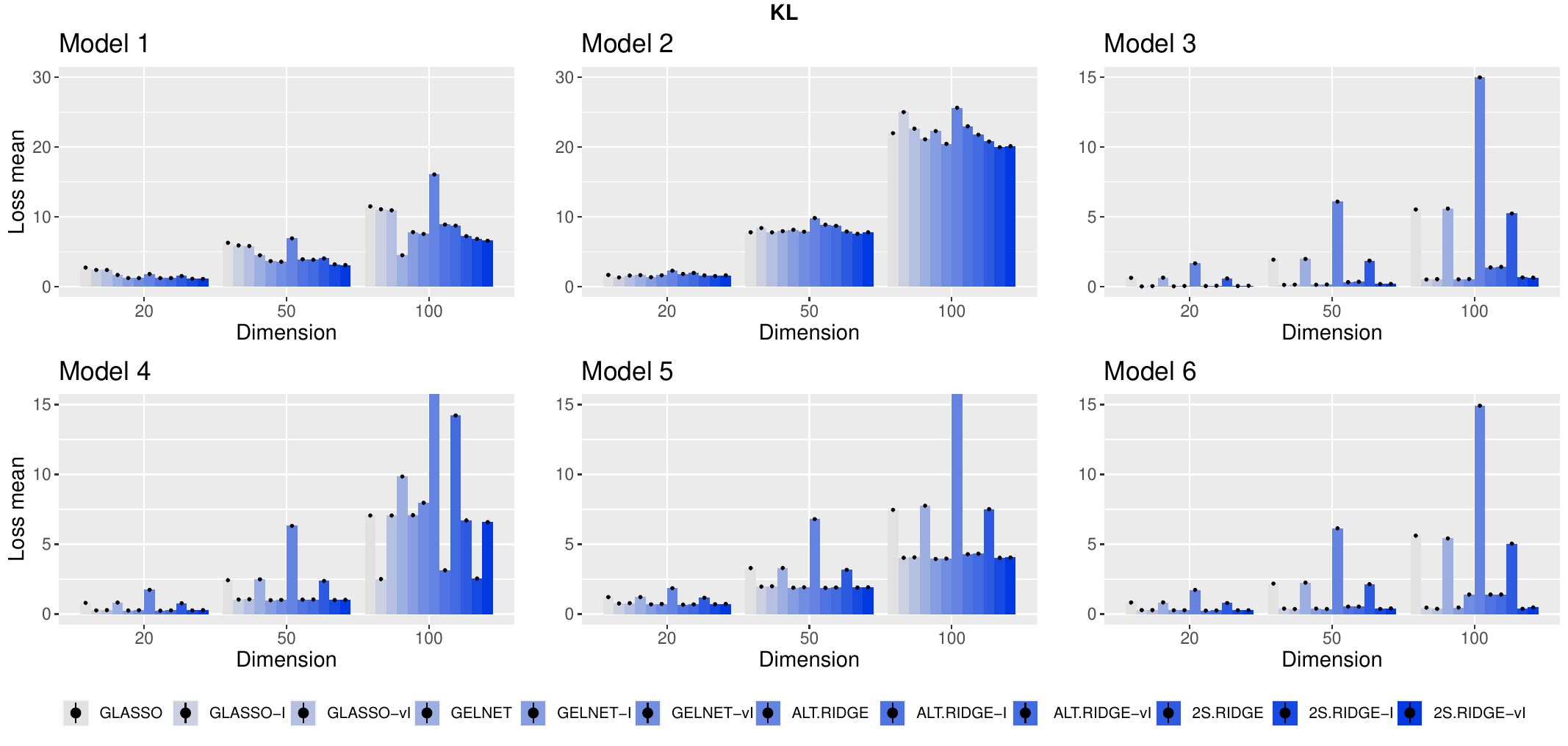}
    \caption{The mean of the Kullback–Leibler loss for different models and methods based
on $100$ replications.}
    \label{fig:mesh01}
\end{figure}

\begin{figure}[h!]
    \centering
    \includegraphics[scale=0.42]{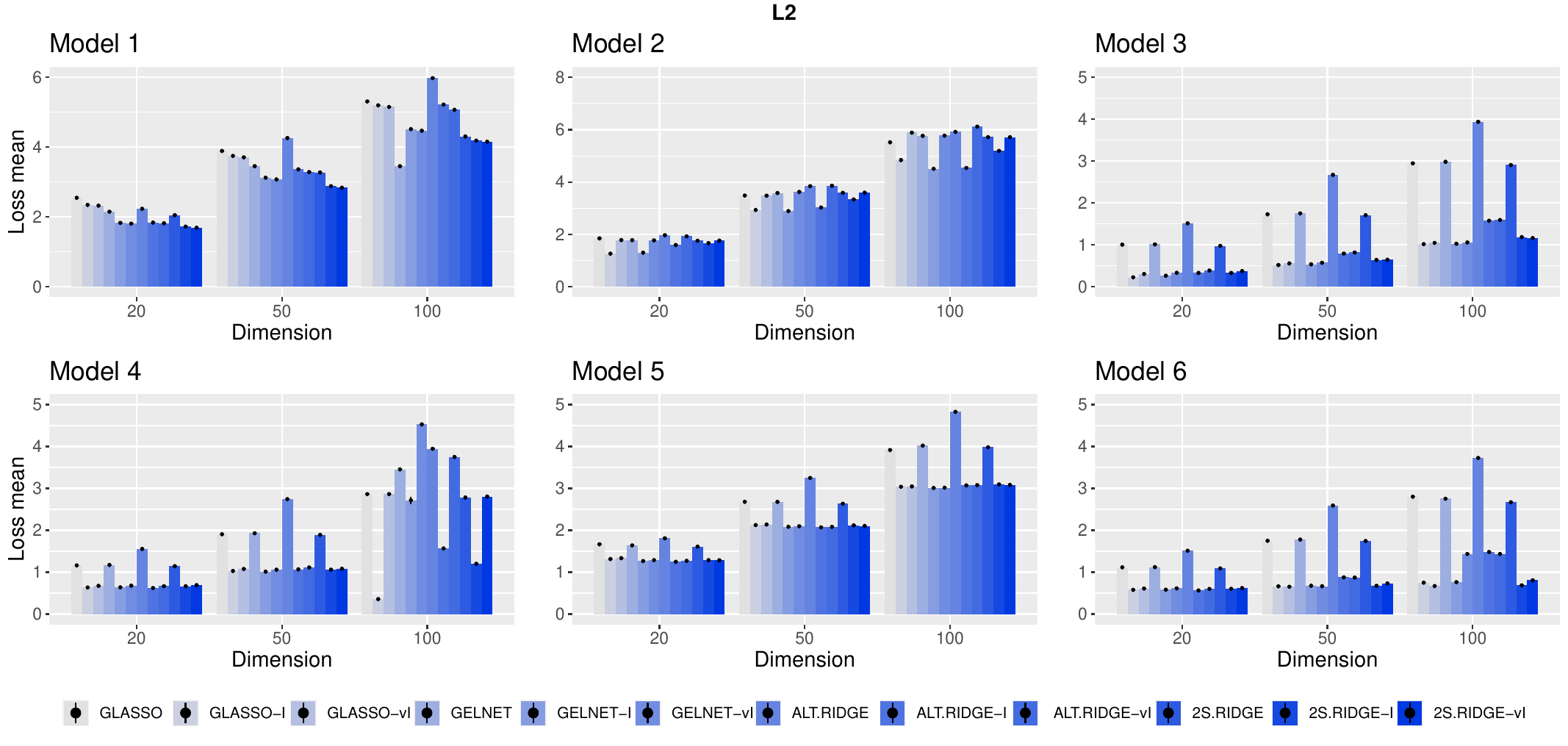}
    \caption{The mean of $L2$ loss for different models and methods based on $100$ replications.}
    \label{fig:mesh02}
\end{figure}

\begin{figure}[h!]
    \centering
    \includegraphics[scale=0.42]{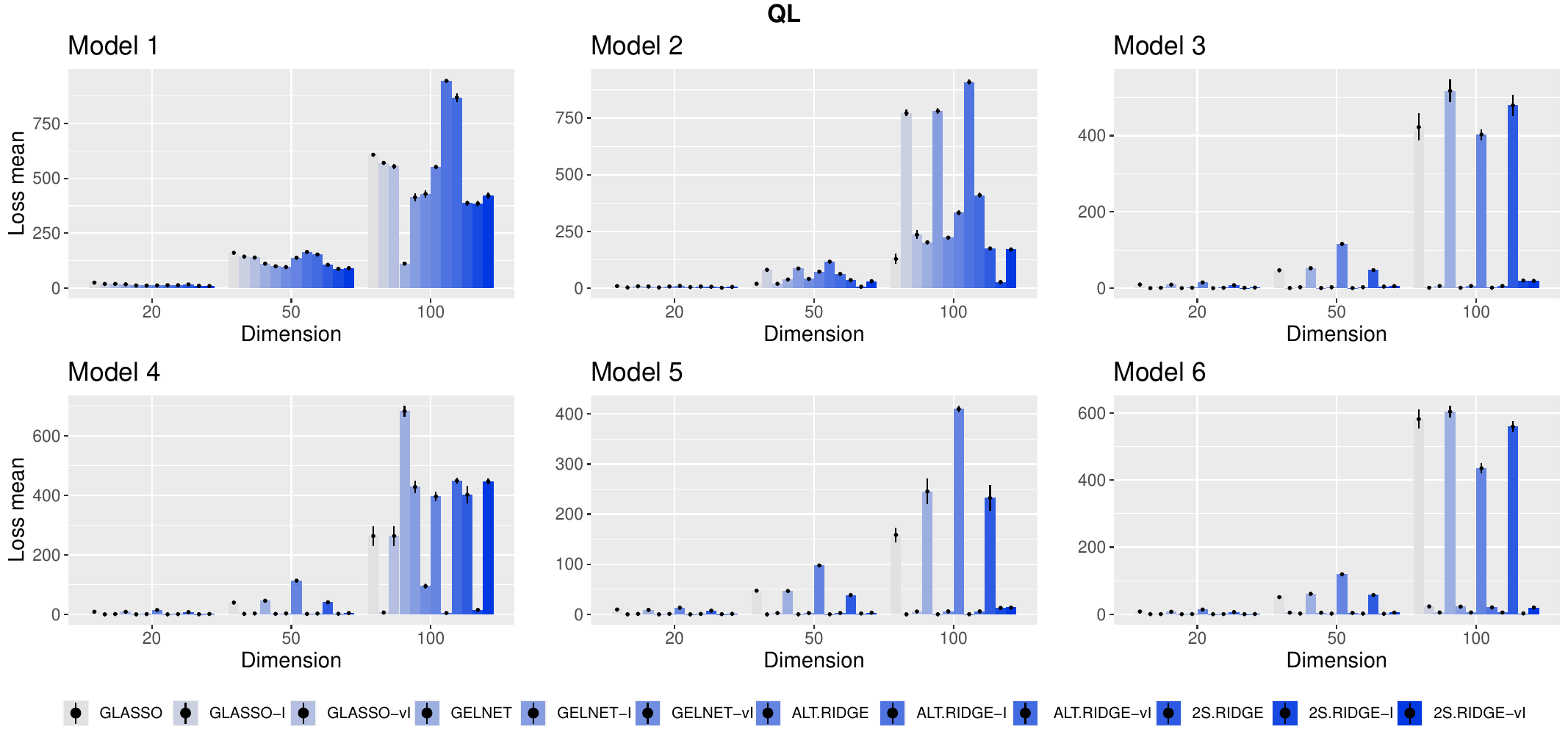}
    \caption{The mean of quadratic loss for different models and methods based on $100$ replications.}
    \label{fig:mesh03}
\end{figure}

\begin{figure}[h!]
    \centering
    \includegraphics[scale=0.42]{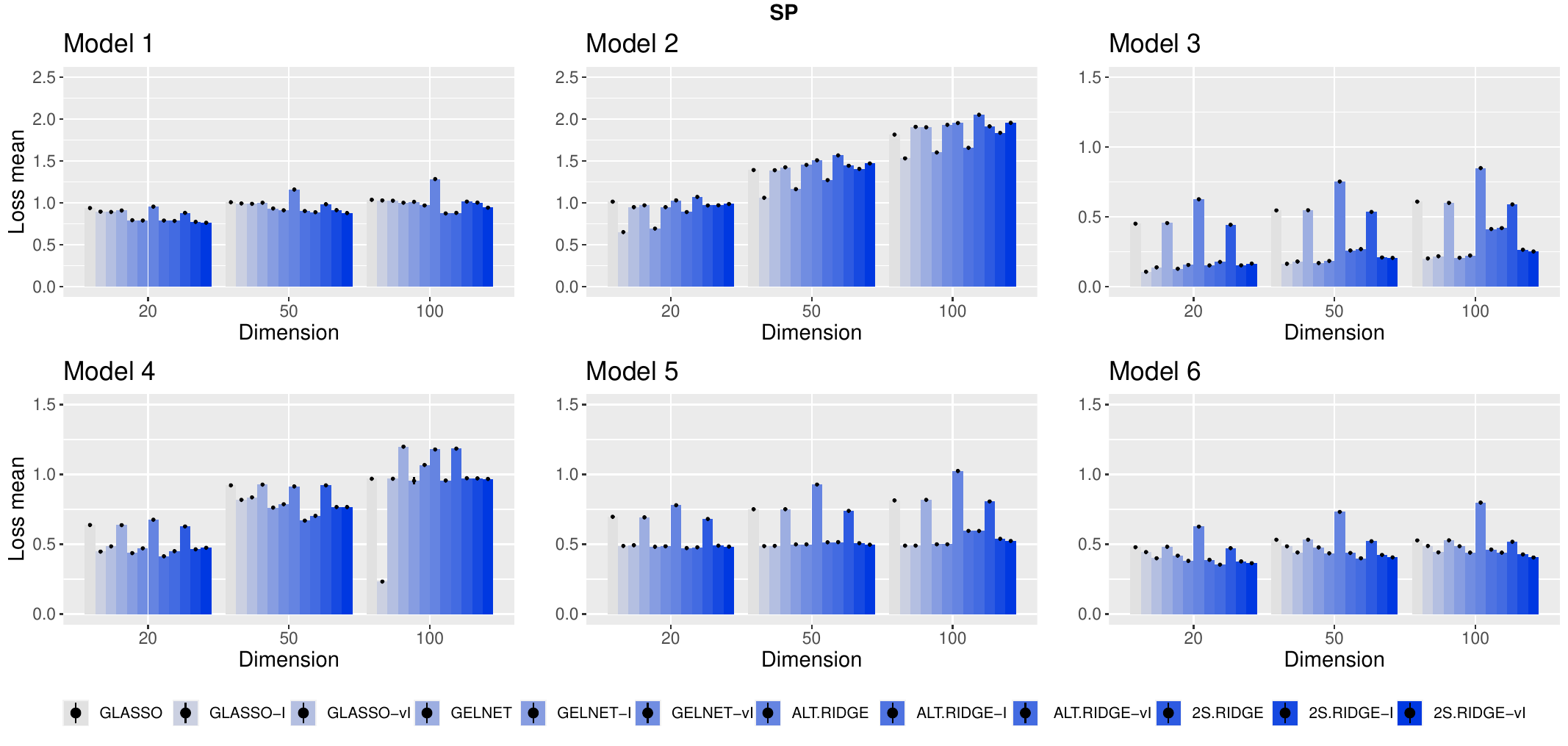}
    \caption{The mean quadratic loss for different models and methods based on 100 replications.
    }
    \label{fig:mesh04}
\end{figure}

 To compare the methods, the averages of the aforementioned loss functions are computed based on $100$ simulations. Figures \ref{fig:mesh01}–\ref{fig:mesh04} present the final results of the evaluation. In these figures, the columns represented by small black dots represent the mean values of the loss functions, while the bars positioned on top of each column indicate the standard errors (shown as mean ± SD). These visual representations provide an overview of the performance of each method, with the mean values indicating the average performance and the error bars providing an indication of the variability or uncertainty associated with the estimates. The plot layout is taken from \cite{kuismin2017precision}.
The results for different networks are displayed in Figures \ref{fig:mesh01}-\ref{fig:mesh04}. We summarize some observations based on the results as follows:
\begin{itemize}
    \item In general, incorporating a target matrix in the estimation methods can be advantageous because each approach tends to perform better when paired with the appropriate target. \\
    \item The Alternative ridge with target performs better than Glasso with no target (as \cite{kuismin2017precision} mentioned before). Also, the 2-step ridge outperforms the Alternative ridge in most considered measures and networks. (in Kullback–Leibler loss, the 2-step ridge shows better results than the alternative ridge in all networks.)\\
    \item  The effectiveness of different target matrices remains a question, but based on our simulations, it is generally observed that the Identity target matrix yields better results in most cases.\\
    \end{itemize}
\section{Application}\label{se5}
In this section, we investigate the performance of our proposed estimator in analysing three
different datasets.
\subsection*{The Ionosphere dataset}
In the present section, our focus is on the ionosphere dataset (available at \url{http://archive.ics.uci.edu/ml/datasets/Ionosphere}). Here, we analyze the performance of linear discriminant analysis (LDA) by computing misclassification rates. This analysis involves the utilization of various estimates for covariance and precision matrices, alongside the implementation of the Mahalanobis distance to partition samples into two distinct groups.\\
The dataset was gathered by a system located in Goose Bay, Labrador, focusing on capturing information about free electrons present in the ionosphere. Radar returns classified as "good" exhibit noticeable structure within the ionosphere, while "bad" returns indicate signals that penetrate through the ionosphere without any evident features. The dataset comprises a total of 351 observations and 35 variables, among which one variable is nominal with the values "good" or "bad," while the remaining 34 variables are continuous.\\
To compare the considered estimators, we adopted the methodology used by \cite{kuismin2017precision} in their study on the impact of covariance matrix and precision matrix estimations on the performance of Linear Discriminant Analysis (LDA) for classification. They demonstrated the effect of different estimators on the classification accuracy of LDA using this dataset. In our current study, we aim to replicate their methodology and apply it to the estimators we are considering. By doing so, we can evaluate how well our proposed estimator performs in terms of its impact on LDA classification accuracy.

According to their work, the dataset consists of two groups: $G_1$ representing good radar returns and $G_2$ representing bad radar returns. The population mean vectors for $G_1$ and $G_2$ are denoted as $\bm\mu_1$ and $\bm\mu_2$, respectively. The precision matrix of the population, denoted as $\bm\Theta$, is assumed to be the inverse of the covariance matrix $\bm\Sigma$. The dataset is divided into a training set and a test set. The training set consists of 40 observations, which is close to the dimension of the data ($p = 32$). The remaining 311 observations are included in the test set. The goal is to estimate $\bm\mu_1$, $\bm\mu_2$, and $\bm\Theta$. The sample means $\overline{\mathbf{x}}_1$ and $\overline{\mathbf{x}}_2$ are used to estimate $\bm\mu_1$ and $\bm\mu_2$, respectively. Different estimators are employed to estimate $\bm\Theta$. Each observation $\mathbf{x}$ in the test set is classified into either $G_1$ or $G_2$ using the LDA rule. The classification is based on the squared Mahalanobis distance, which is computed as $\mathbf{a}^T(\mathbf{x}-\bm\mu)$, where $\mathbf{a}=\bm\Theta(\bm\mu_1-\bm\mu_2)$ and $\bm\mu=0.5(\bm\mu_1+\bm\mu_2)$. The misclassification rate is estimated by applying the classification rule obtained from the training set to the observations in the test set.

In our study, we consider Glasso, Gelnet, and Alternative ridge as alternative estimators for the precision matrix $\Theta$. These estimators are used to estimate $\Theta$, which is then utilized to allocate an observation $\mathbf{x}$ to either group $G_1$ or $G_2$. To determine the group allocation, we calculate the estimated discriminant $\hat{\mathbf{a}}^T(\mathbf{x}-\hat{\bm\mu})$, where $\hat{\mathbf{a}}=\hat{\Theta}(\overline{\mathbf{x}}_1-\overline{\mathbf{x}}_2)$ and $\hat{\bm\mu}=0.5(\overline{\mathbf{x}}_1+\overline{\mathbf{x}}_2)$. If the calculated discriminant value is greater than zero, the observation $\mathbf{x}$ is assigned to group $G_1$. 

By employing these alternative estimators for $\Theta$, we aim to assess how they influence the classification performance of the LDA model. This approach allows us to compare the performance of the estimators and determine their effectiveness in accurately allocating observations to the appropriate groups.
As target matrices, we used a zero matrix, identity matrix $I$ and a scalar matrix $\nu I$, where $\nu=p^2/tr(S)$. The associated misclassification rates are the mean values obtained from 100 iterations of the previously indicated setting. 

In Figure \ref{fig:Name:a}, we present the misclassification rates obtained for all the estimators considered in our study, across different values of the tuning parameter. The tuning parameter, denoted as $\rho$, is varied within a series of length 50, ranging from $0.1$ to $0.9$. For the estimators Glasso and Alternative ridge, the tuning parameter $\rho$ is directly applied. As for Gelnet and 2-step ridge, we set $\alpha$ to $0.2$, and the value of the tuning parameter $\lambda$ is set equal to $\rho$.

For further analysis, the data is randomly split into three sets: a training set with $40$ observations, a validation set with $40$ more observations, and a test set with the remaining observations. Cross-validation for Glasso, Gelnet, Alternative ridge, and the 2-step ridge is used to choose the tuning parameter from the validation set. The precision matrix estimates are then derived using this tuning parameter value from the training set. Finally, we compute the misclassification errors using the test set as our basis. This process is repeated $100$ times. Figure \ref{fig:Name:b} displays the boxplots of the $100$ misclassification errors made using this approach.
\begin{figure}[h!]
\centering
\begin{tabular}{c}
\includegraphics[width=0.5\textwidth]{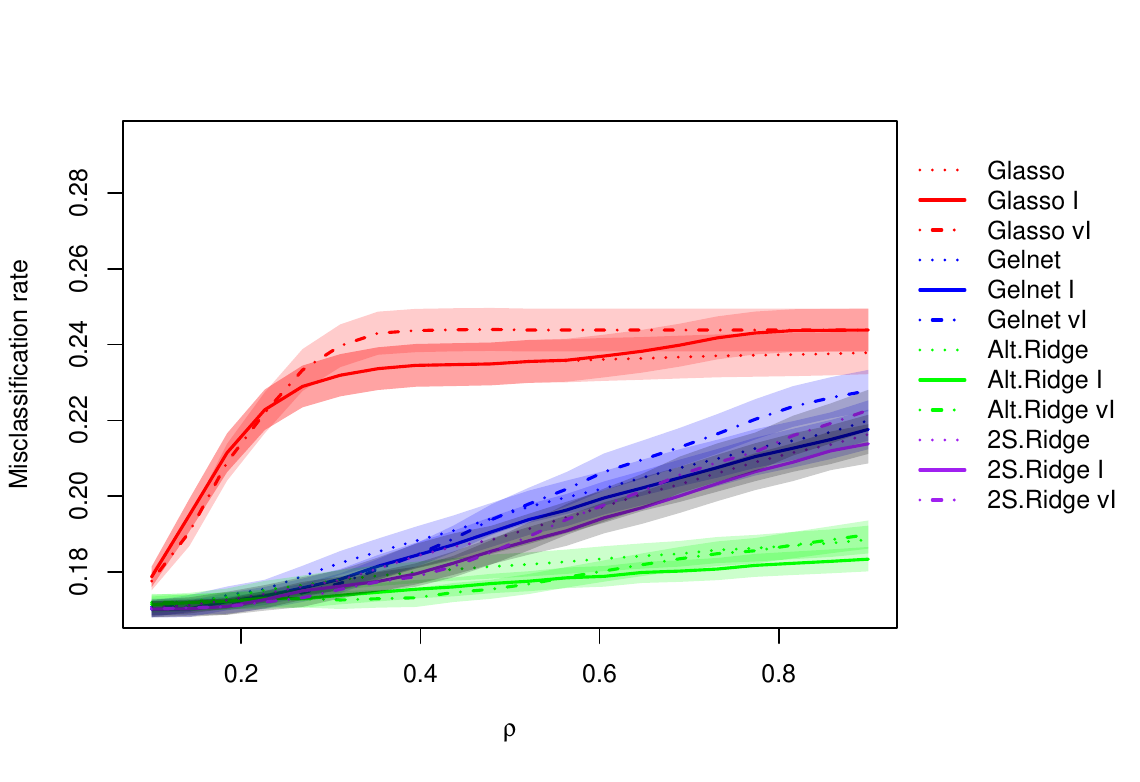} \\
\footnotesize\textbf{(a)} \\
\includegraphics[width=0.5\textwidth]{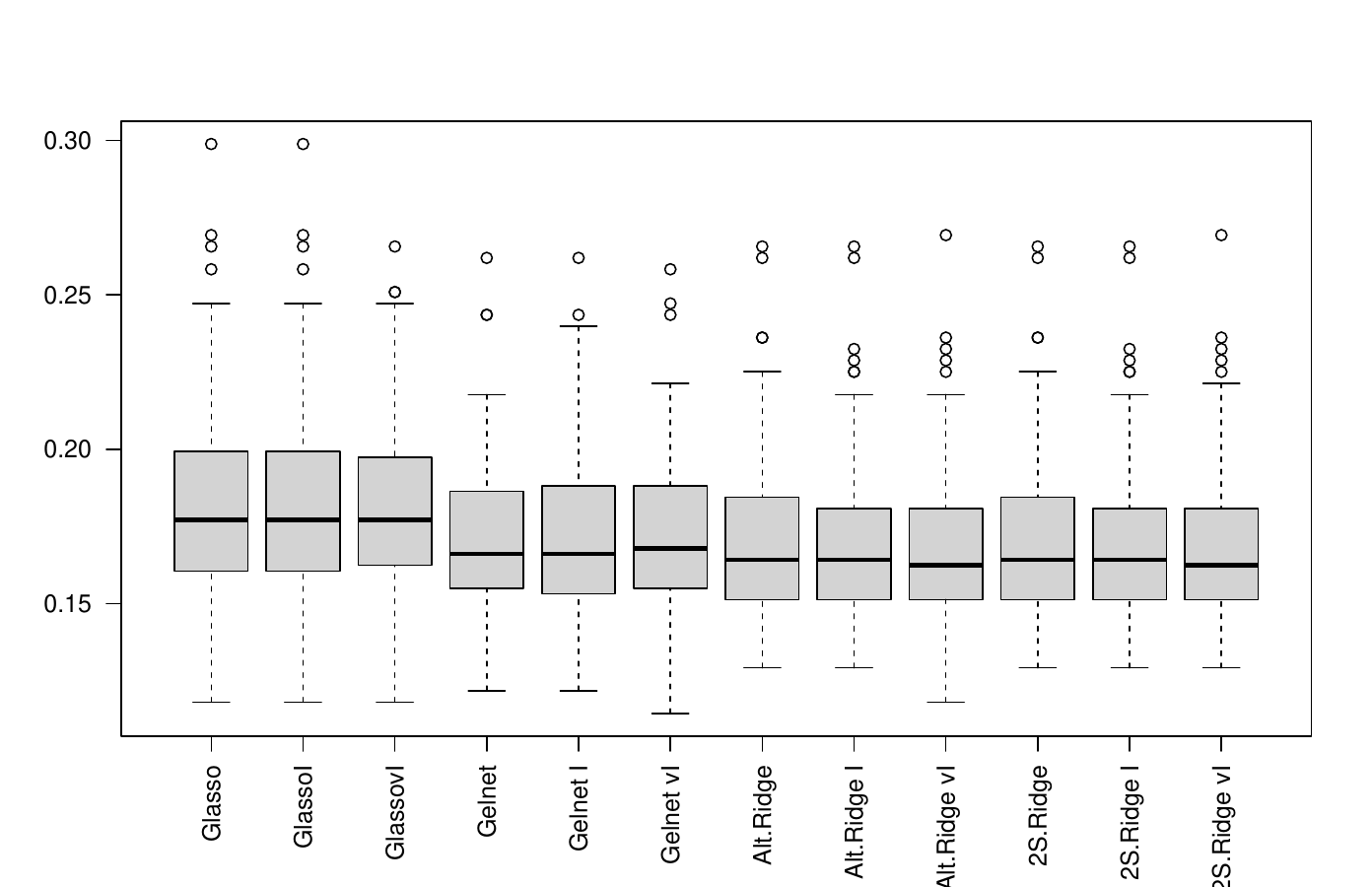} \\
 \\ \footnotesize\textbf{(b)} 
\end{tabular}
\caption{ Ionosphere dataset: \textbf{(a)} The misclassification rates of LDA and standard errors (shaded areas)
\textbf{(b)} boxplots of test misclassification errors
}
\label{fig:Name}
\extralabel{fig:Name:a}{(a)}
\extralabel{fig:Name:b}{(b)}
\end{figure}
For the small value of $\rho$, 2-step ridge, Gelnet, and Alternative ridge behave similarly and give a lower misclassification rate than Glasso, but when $\rho$ increases, the Alternative ridge gives a lower rate than 2-step ridge, and 2-step ridge gives a lower rate than Gelnet. The 2-step ridge has the smallest standard error, slightly different from the Alternative ridge.
When the tuning parameter is chosen with cross-validation from the validation set (\ref{fig:Name:b}), 2-step ridge and Alternative ridge give the lowest median calculated for the misclassification error: $0.164$ (zero target matrix), $0.164$ (Identity target matrix), $0.162$ (scalar target matrix). In this example, the Alternative ridge estimator gives a lower misclassification rate; considering a small value for tuning parameter $\alpha$, the 2-step ridge can get close to this rate. Now we consider another real example and show that our proposed estimator has a lower misclassification rate. 
\subsection*{Wisconsin breast cancer dataset}
This dataset is available at \url{https://archive.ics.uci.edu/ml/datasets/breast+cancer+wisconsin+(diagnostic)} and 
contains 569 observations and 31 variables, including
one nominal variable with the values ”malignant” or ”benign” and 30 continuous variables. The problem is distinguishing cancerous, malignant, from non-cancerous, benign. We follow the same strategy previously to compare the misclassification rate for Alternative ridge and 2-step ridge estimators. 
\begin{figure}[h!]
\centering
\begin{tabular}{c}
\includegraphics[width=0.5\textwidth]{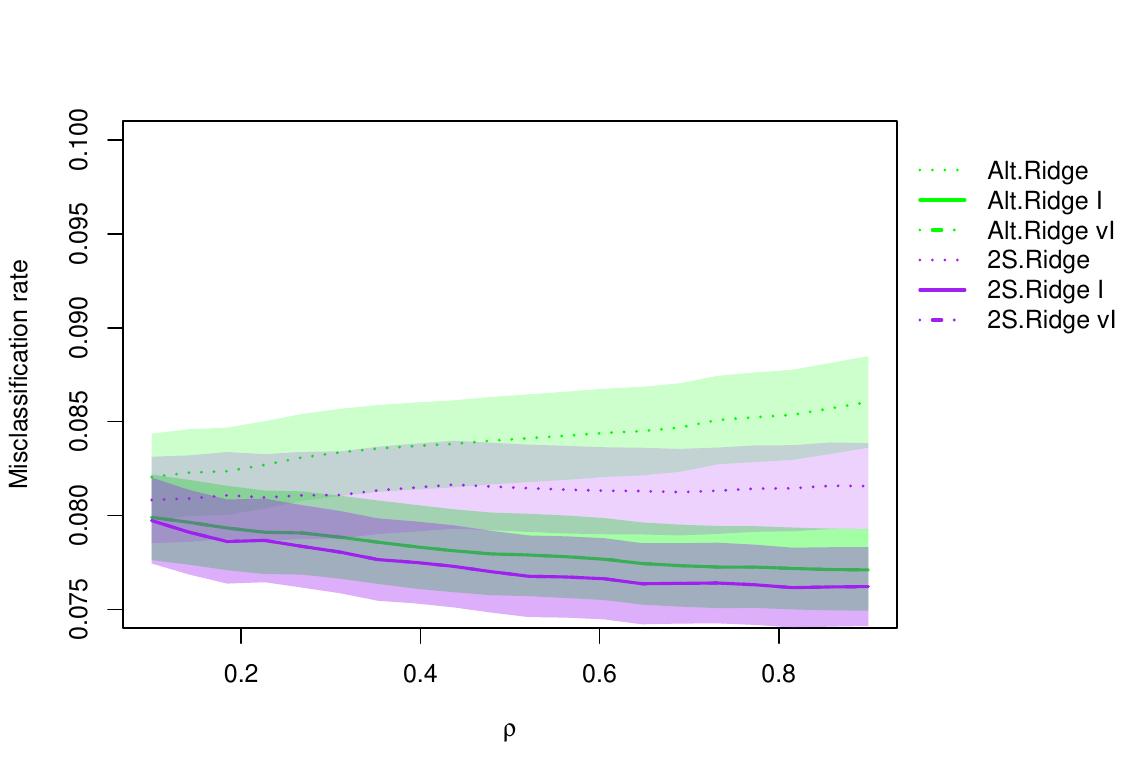}\\
\footnotesize\textbf{(a)} \\
\includegraphics[width=0.5\textwidth]{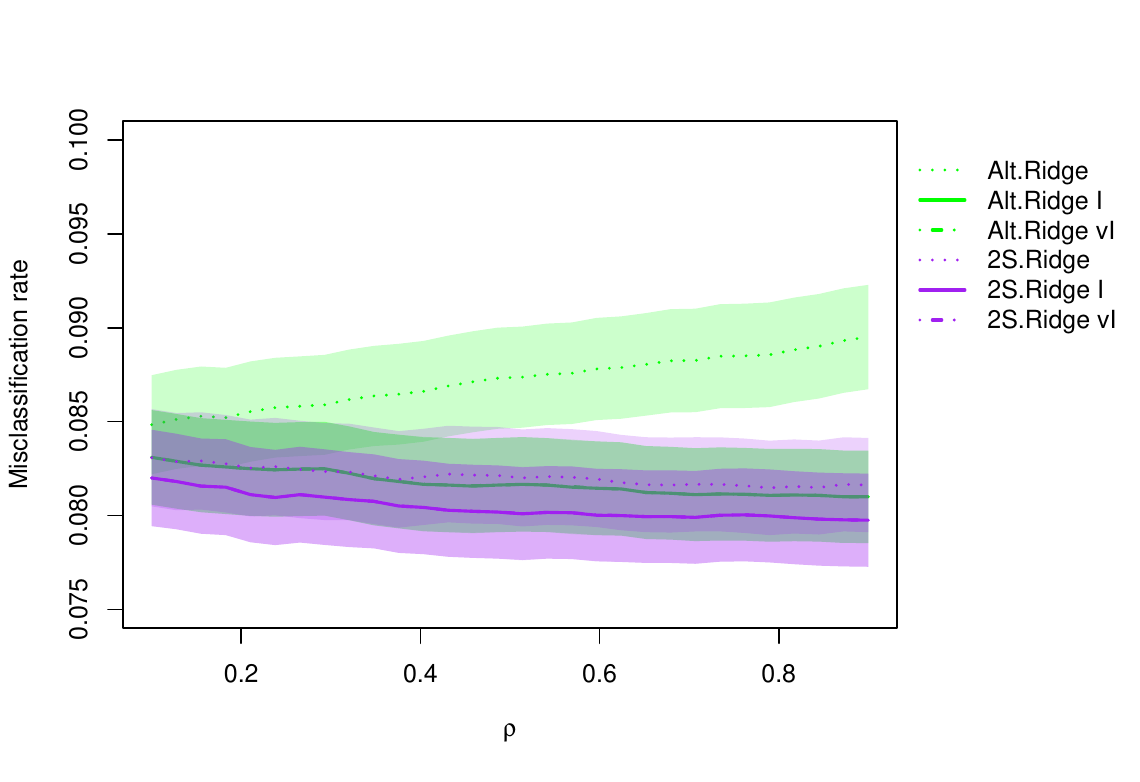}\\
\footnotesize\textbf{(b)} 
\end{tabular}
\caption{Wisconsin breast cancer dataset: The misclassification rates of LDA and standard errors (shaded areas)
when
\textbf{(a)} $\alpha=0.2$.
\textbf{(b)} $\alpha=0.5$.
}
\label{fig:mesh06}
\extralabel{fig:mesh06:a}{(a)}
\extralabel{fig:mesh06:b}{(b)}
\end{figure}
Figure \ref{fig:mesh06:a} shows the 
misclassification rates for Alternative ridge for each tuning parameter value $\rho$ from a series of length $50$ ranging from $0.1$ to $0.9$ and 2-step ridge for $\alpha=0.2$ and $\lambda=\rho$. Figure \ref{fig:mesh06:b} displays misclassification rates when $\alpha=0.5$.
Figure \ref{fig:mesh06} shows the misclassification rates of the 2-step ridge are lower than the Alternative ridge estimators in this example for both consider tunning parameter $\alpha$.\\
\subsection*{Top 100 cryptocurrencies historical dataset }
Here, we consider the top 100 cryptocurrencies' historical dataset, which is available on Kaggle. The dataset contains the open, high, low, close and volume values of the cryptocurrencies priced based on USD. We examine a total of 30 cryptocurrencies over a span of three years, specifically from 2019 to 2021. The names and corresponding abbreviations of these 30 cryptocurrencies can be found in Table \ref{tab1}. 
\begin{table}
\caption{List of abbreviations of the considered cryptocurrencies.}\label{tab1}
\vspace{0.3cm}
\footnotesize{
\begin{tabular}{llllll}
\hline
\multicolumn{4}{l}{Cryptocurrencies}
\\ \hline
ADA  & Cardano & EOS  & Electro-Optical System & NEM   & New Economy Movement\\
BAT  & Basic Attention Token & ETC  & Ethereum Classic & QTUM  & Qtum\\
BCH  & Bitcoin Cash & ETH  & Ethereum & TRX   & TRON\\
BNB  & Binance Coin &  FIL  & Filecoin & USDT  & Tether\\
BTC  & Bitcoin & GNO  & Gnosis  & WAVES & WAVES\\
BTG  & Bitcoin Gold & IOTA  & IOTA  & XLM & Stellar\\
DASH & Digital Cash & KCS   & KuCoin Shares & XMR   & Monero\\
DCR  & Decred  & LRC & Loopring & XRP   & Ripple\\
DOGE & Dogecoin  & LTC   & Litecoin & XTZ   & Tezos\\
ENJ  & Enigma  & MANA  & Decentraland & ZEC   & ZCash 
\end{tabular}}
\end{table}
We compute the daily log returns of the closed prices for the remaining 30 cryptocurrencies. 

\begin{figure}
\centering
\begin{tabular}{cc}
\includegraphics[width=0.5\textwidth]{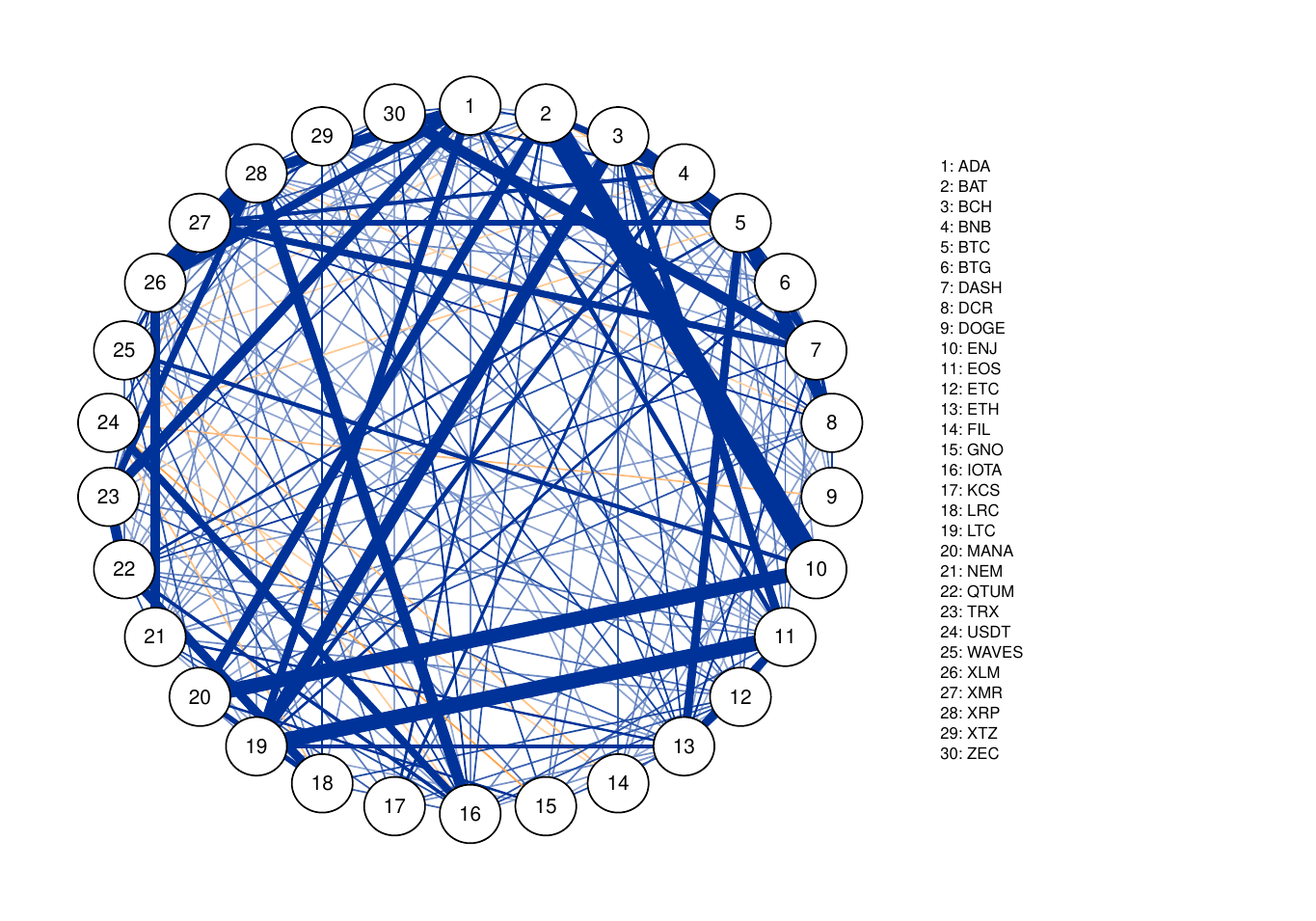} & \includegraphics[width=0.5\textwidth]{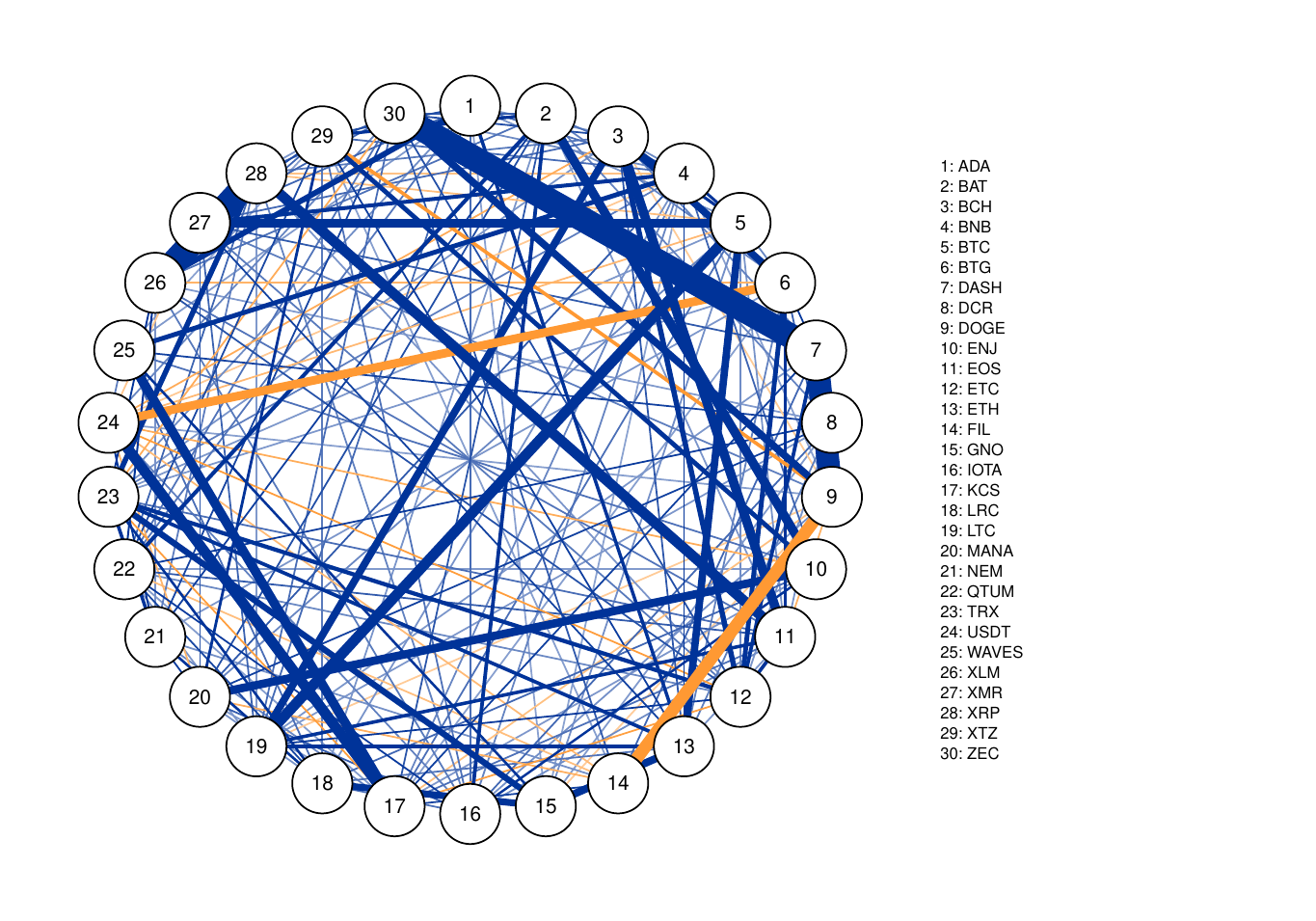}\\
\footnotesize\textbf{(a)} Estimated network based on 2019 & \footnotesize\textbf{(b)} Estimated network based on 2020\\
\includegraphics[width=0.5\textwidth]{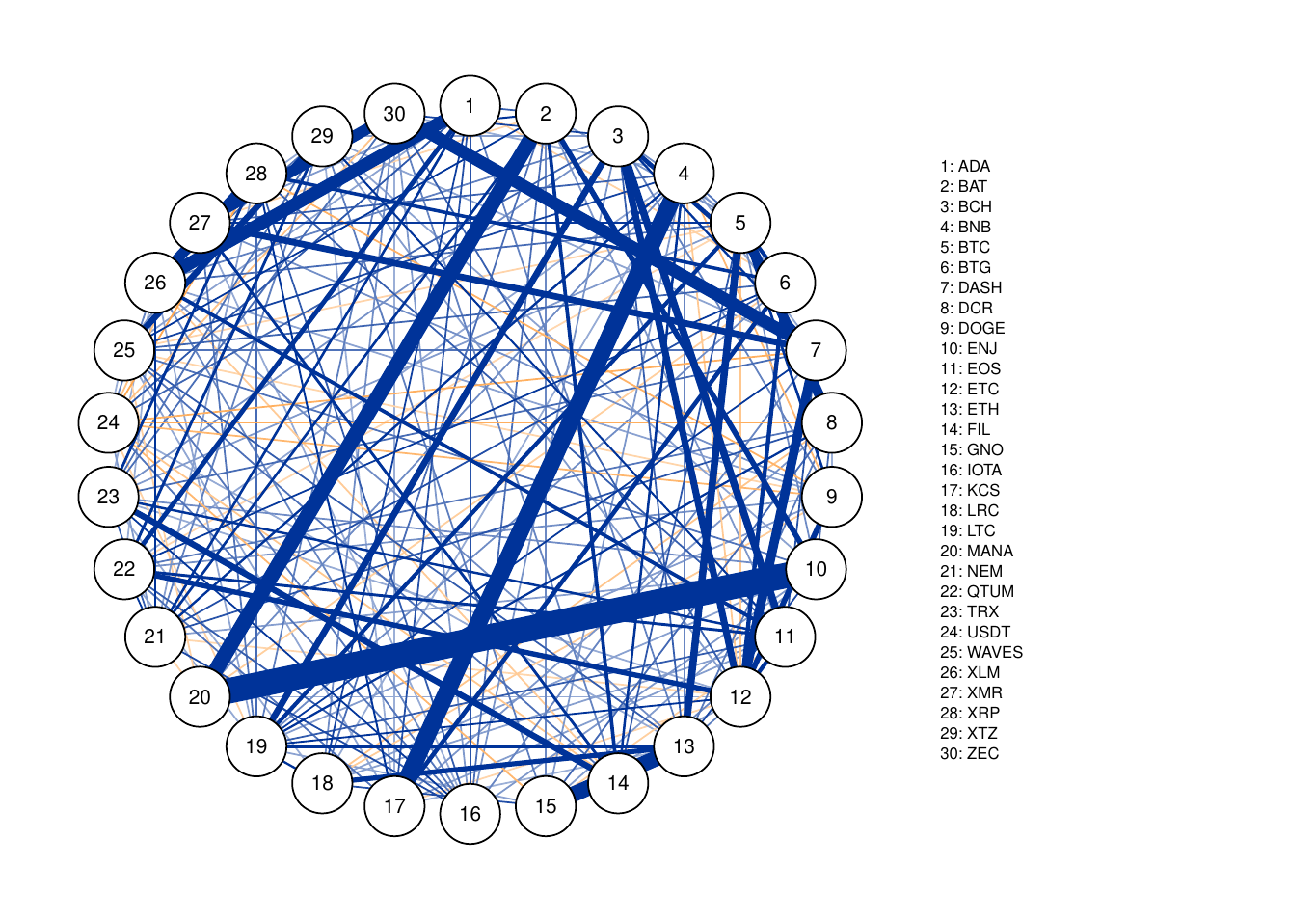} & \includegraphics[width=0.5\textwidth]{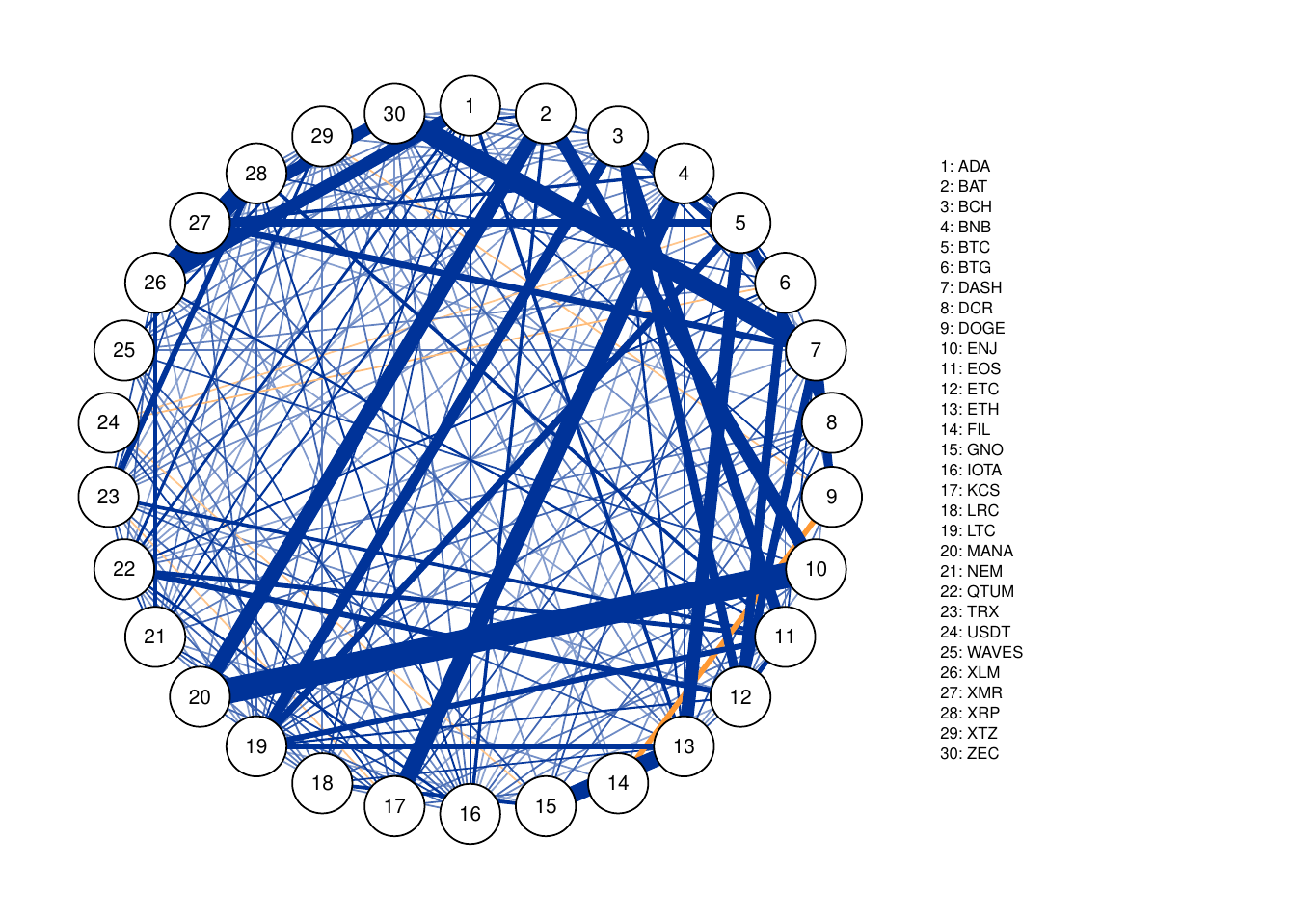} \\
 \\ \footnotesize\textbf{(c)} Estimated network based on 2021 & \footnotesize\textbf{(d)} Estimated network based on 2019-2021
\end{tabular}
\caption{Estimated cryptocurrency networks based on 2019 to 2021 using 2-step ridge.}
\label{fig:7}
\extralabel{fig:7:a}{(a)}
\extralabel{fig:7:b}{(b)}
\extralabel{fig:7:c}{(c)}
\extralabel{fig:7:d}{(d)}
\end{figure}

We utilize our proposed estimator to estimate the networks using both three-year and annual data separately. In the estimation process, the Identity matrix is used as the target. To select the tuning parameters, we employ five-fold cross-validation. The estimated networks obtained from these estimations are presented in Figure \ref{fig:7}. The edges that correspond to negative partial correlations are indicated in orange, and the edges that belong to positive partial correlations are indicated in blue.
The width of the edge in the network visualization is proportional to the magnitude of the partial correlation estimates.  
\begin{figure}[h!]
\centering
\begin{tabular}{cc}
\includegraphics[width=0.5\textwidth]{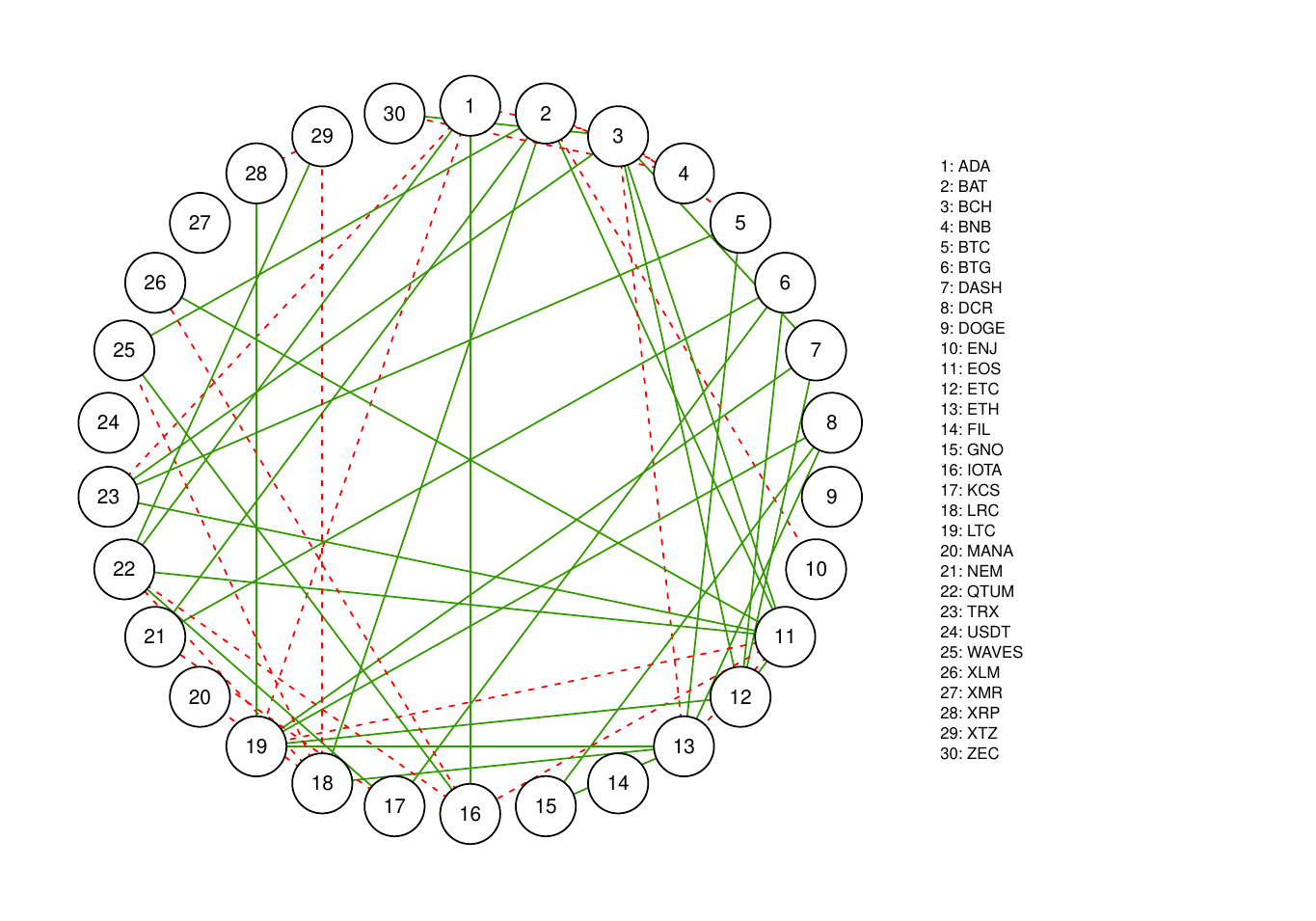} & \includegraphics[width=0.5\textwidth]{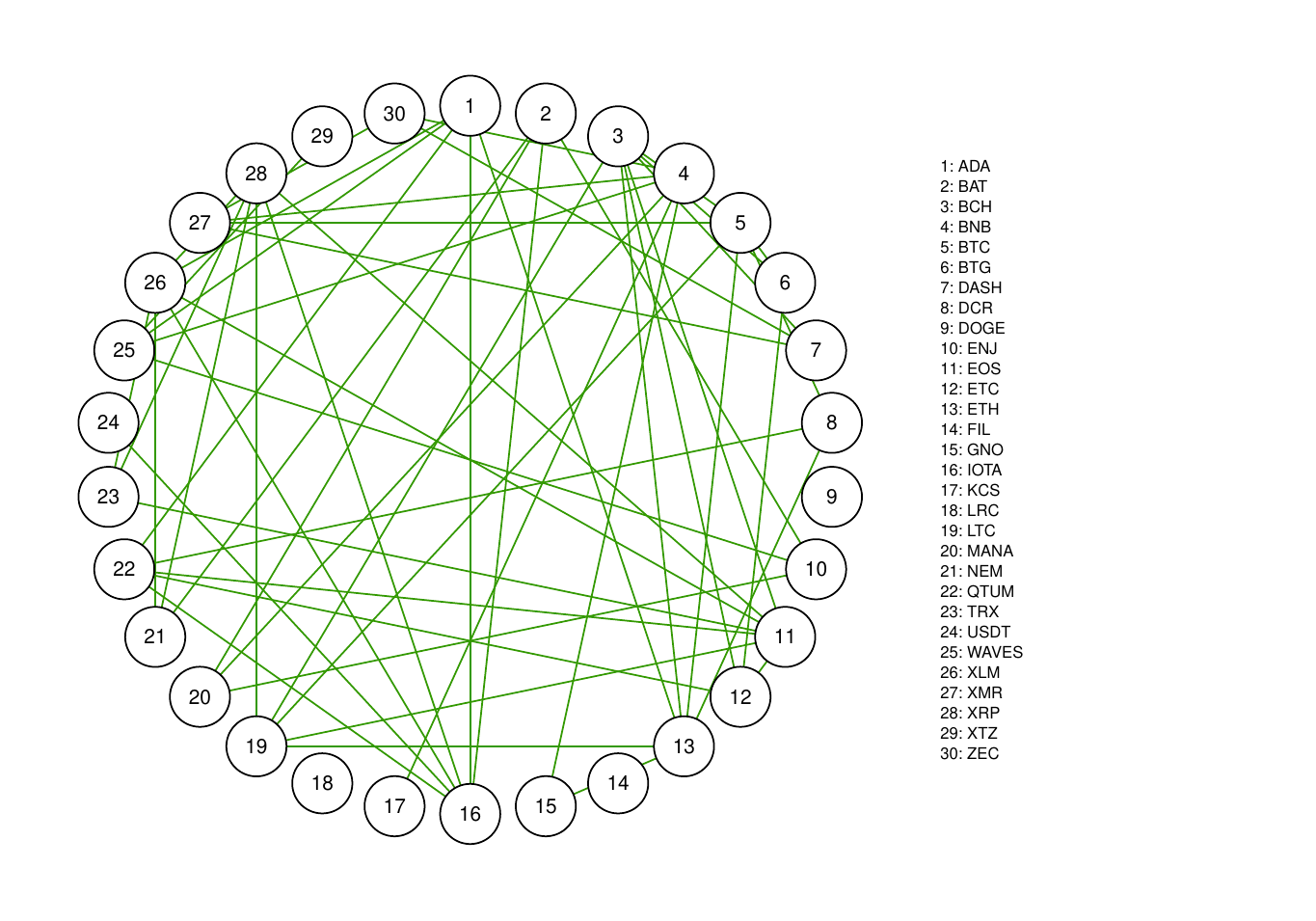}\\
\footnotesize\textbf{(a)} Ascending-descending trend of partial correlations & \footnotesize\textbf{(b)} Stable relations based on 2019-2021
\end{tabular}
\caption{Ascending-descending trend and Stable relations.}
\label{fig:8}
\extralabel{fig:8:a}{(a)}
\extralabel{fig:8:b}{(b)}
\end{figure}

\begin{figure}[h!]
    \centering
    \includegraphics[scale=0.38]{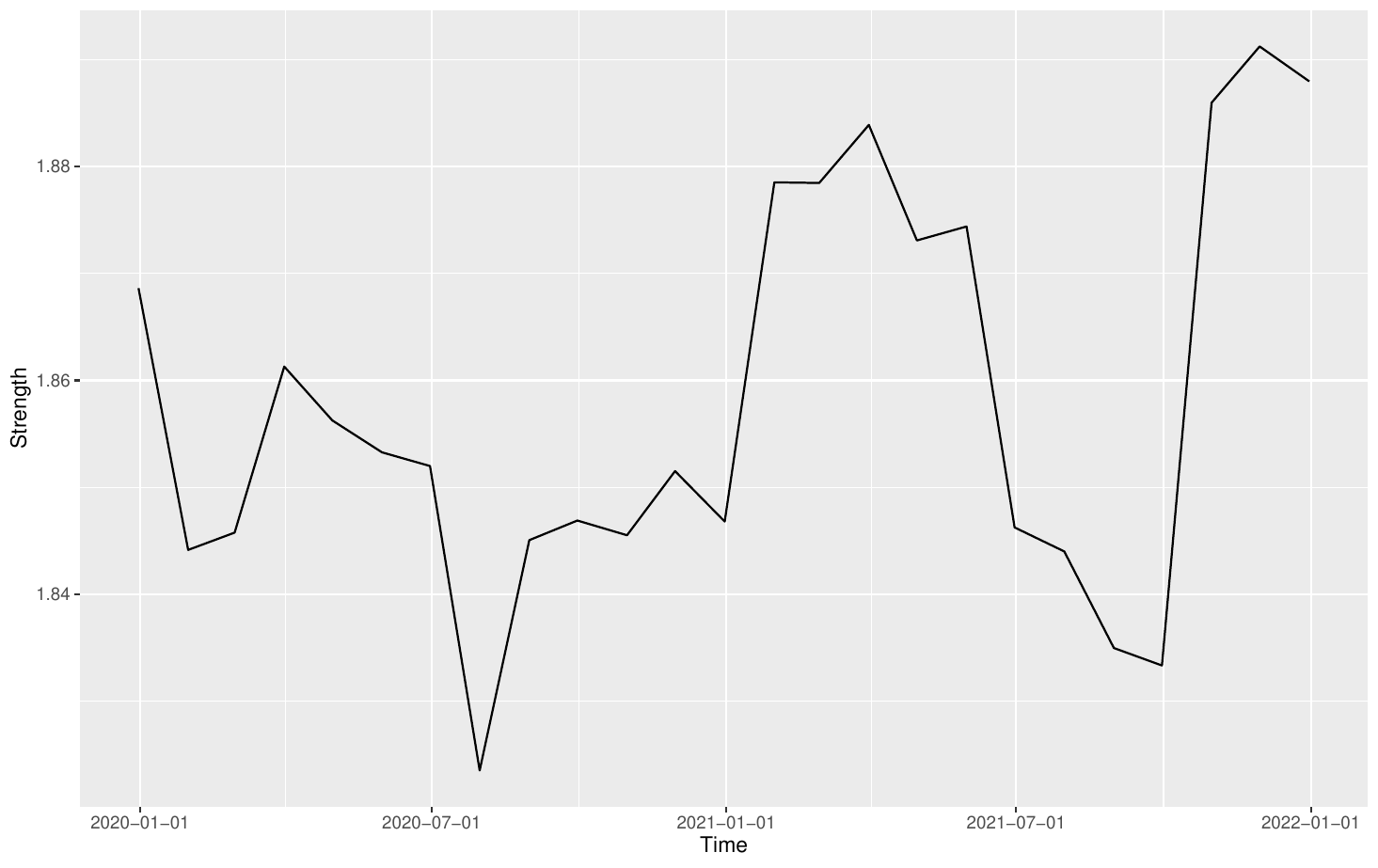}
    \caption{The mean value of strength estimated on a rolling window of one year with shifts of one month.}
    \label{fig9}
\end{figure}

 Figure \ref{fig:8:a} shows the increasing (in green) and decreasing (in red) trend of partial correlations. Figure \ref{fig:8:b} shows the stable relationships between all dependencies based on a three-year estimate.
So, for instance, the partial correlation between Bitcoin and Ethereum is increasing across 2019, 2020, and 2021, while the partial correlation between Bitcoin and Bitcoin Cash is decreasing during the same time. Also, about \ref{fig:8:b}, for example, the connection between Bitcoin and Monero is stable across 2019, 2020 and 2021, while the connection
between Bitcoin and Digital Cash is not stable because it is only detected in 2019.\\
As a network measure, we provide the mean strength value.
The strength of a node is equal to the sum of the weights (partial correlations) of all connected edges. This measure shows the intensity or strength of the connections between currencies. Figure \ref{fig9} presents the mean value of the strength estimated using a rolling window of one year, with shifts of one month. This rolling window analysis allows us to capture the changes in strength over time and observe any patterns or trends in the data.
We observe a decrease in the mean values of strength between 2019 and 2020, followed by an increase in the first month of 2021 and a more noticeable increase in the latest months of 2021. The high mean strength values in 2021 can explain the similarity of the estimated network 2021 to the three-year estimated network.
\section{Conclusions}\label{sec6}
Including a reasonable target matrix seems to improve the estimation. It is also mentioned by \cite{kovacs2021graphical}. Furthermore, we can not choose one estimator as the best. According to the simulation study and real data examples, the better performance of the 2-step ridge precision estimator is visible than the alternative ridge estimator with corresponding target matrices. However, the alternative ridge is superior to our suggested estimate in terms of computational time. Finally, We consider two real data examples to compare the proposed estimator to the alternative ridge. Followed by an empirical examination of the network across 30 cryptocurrencies. We consider the daily closed prices and the 2-step ridge to estimate their interactions and, in turn, the perceived network.

\section*{Acknowledgement}
This work was based upon research supported in part by the National Research Foundation of South Africa Ref.: SRUG2204203965, and RA171022270376, UID 119109; the Department of Research and Innovation at the University of Pretoria (SA), STATOMET, as well as the Centre of Excellence in Mathematical and Statistical Sciences, based at the University of the Witwatersrand (SA).  The opinions expressed and conclusions arrived at are those of the authors and are not necessarily to be attributed to the NRF. The second author's research (M. Arashi) is supported by a grant from Ferdowsi University of Mashhad (N.2/58091).

\bibliographystyle{tfs}
\bibliography{interacttfssample}
\clearpage
\section*{Appendix A: Proofs}
\textit{\textbf{Lemma A1}(Theorem 3.1 of \cite{shirilord2022closed})}: 
Let $\bm A,\bm B, \bm C, \bm D$ be real or complex value $n\times n$ matrices. Suppose that $\bm C$ is nonsingular and the matrices $\bm A, \bm B$ and $\bm C$
satisfy $\bm B\bm C = \bm C\bm A$. Denote $\bm\Delta:=\bm A^2+\bm D\bm C$, be a nonsingular matrix. Then a closed-form expression for the solution of a non-symmetric algebraic Riccati equation, $\bm A\bm X+\bm X\bm B+\bm X\bm C\bm X=\bm D$ is given by the following formula 
\begin{equation}
    \bm X=(\sqrt{\bm\Delta} -\bm A)\bm C^{-1},
\end{equation}
where $\sqrt{\bm M}$ is square root of matrix $\bm M$.
\\\\
\textbf{\textit{Proof of Proposition \ref{prop1}}}:
Consider the dual optimization of problem \eqref{e01} by exchanging the max and the min. The resulting inner problem in $\bm \Theta$ can be solved analytically by setting the gradient of the objective to zero and solving for $\bm\Theta$, we have
\begin{align}\label{e002}
\nonumber    &\bm\Theta^{-1}-\bm S-\bm U+\lambda(1-\alpha)(\bm \Theta-\bm T)=\bm 0,\\
    &\bm\Theta^{-1}-\bm S-\bm U+\lambda(1-\alpha)\bm T-\lambda(1-\alpha)\bm \Theta=\bm 0.
\end{align}
Multiplying the LHS of \eqref{e002} by $\bm \Theta$, we get
\begin{equation}\label{e02}
(\bm S+\bm U-\lambda(1-\alpha)\bm T)\bm\Theta+\lambda(1-\alpha)\bm\Theta^2=\bm I_p.
\end{equation}
We use Lemma A1 to solve the quadratic matrix equation \eqref{e02}, therefore we consider the matrices as follows; $\bm A=\bm B=\frac{1}{2}(\bm S+\bm U-\lambda(1-\alpha)\bm T)$, $\bm C=\lambda(1-\alpha)\bm I$ and $\bm D=\bm I$, we have
\begin{equation}
    \bm \Delta=(\frac{1}{2}(\bm S+\bm U-\lambda(1-\alpha)\bm T))^2+\lambda(1-\alpha)\bm I.
\end{equation}
Since $\bm C$ and $\bm \Delta$ are non-singular and $\bm B\bm C=\bm C\bm A$ then the solution of equation \eqref{e02} is given by following formula
\begin{equation}\label{e04}
    \bm \Theta^*=\frac{1}{\lambda(1-\alpha)}\left[\left(\mathcal{\bm A}^2+\lambda(1-\alpha)\bm I\right)^{1/2}-\mathcal{\bm A})\right],
\end{equation}
where $\mathcal{\bm A}=\frac{1}{2}(\bm S+\bm U-\lambda(1-\alpha)\bm T)$. Subsequently, we can rewrite \eqref{e01} as
\begin{equation}\label{e05}
    \underset{\parallel \bm U\parallel 	_\infty\leq \lambda\alpha}{min}\log\det (\bm \Theta^*)-tr(\bm \Theta^*,\bm S+\bm U)+\frac{(1-\alpha)}{2}||\bm \Theta^*-T||_2^2.
\end{equation}
The second and third parts of \eqref{e05} can be expressed of the following form
\begin{eqnarray}\label{e06}
\nonumber    tr(\bm \Theta^*(\bm S+\bm U))&+&\frac{(1-\alpha)}{2}||\bm \Theta^*-\bm T||_2^2\\
 &=&tr(\bm \Theta^*(2\mathcal{\bm A}+\frac{\lambda(1-\alpha)}{2}\bm \Theta^*))+\frac{\lambda(1-\alpha)}{2}tr(\bm T^{T}\bm T),
\end{eqnarray}
with substituting \eqref{e04} in \eqref{e06} and doing some calculations, we have
\begin{eqnarray}
\tr(\bm \Theta^*(2\mathcal{\bm A}+\frac{\lambda(1-\alpha)}{2}\bm \Theta^*))=\frac{1}{\lambda(1-\alpha)}tr((\mathcal{\bm A}^2+\lambda(1-\alpha)\bm I)^{1/2}-\mathcal{\bm A})\mathcal{\bm A})+\frac{p}{2}.
\end{eqnarray}
Therefore the optimization problem \eqref{e05} is equivalent to 
\begin{eqnarray}
\nonumber\underset{\parallel \bm U\parallel 	_\infty\leq \lambda\alpha}{\min}\log\det \left(\frac{1}{\lambda(1-\alpha)}\left[\left(\mathcal{\bm A}^2+\lambda(1-\alpha)\bm I\right)^{1/2}-\mathcal{\bm A}\right]\right)\\
\nonumber-\tr\left(\frac{1}{\lambda(1-\alpha)}\left[\left(\mathcal{\bm A}^2+\lambda(1-\alpha)\bm I\right)^{1/2}-\mathcal{\bm A}\right]\mathcal{\bm A}\right).
\end{eqnarray}
which completes the proof.\\\\
\textbf{\textit{Proof of Proposition \ref{prop2}}}:
Since $\bm S+\bm U=\bm P\bm D\bm P^{-1}$ and $\bm T=\gamma \bm I_p$ we can rewrite $\mathcal{\bm A}$ in equation \eqref{e00002} as $\mathcal{\bm A}=\bm P[\bm D-\lambda(1-\alpha)\gamma \bm I_p]\bm P^{-1}$. After some calculations, we have
\begin{eqnarray}
\nonumber\underset{\parallel \bm U\parallel _\infty\leq \lambda\alpha}{\max}  
\tr\left(\bm P\left[\log(\sqrt{\bm B^2+\lambda(1-\alpha)\bm I}+\bm B)+
\frac{1}{\lambda(1-\alpha)}\bm B(\sqrt{\bm B^2+\lambda(1-\alpha)\bm I}-\bm B)\right]\bm P^{-1}\right) ,
\end{eqnarray}
where $\bm B=1/2(\bm D-\lambda(1-\alpha)\bm T)$ and according to this fact that the sum of the eigenvalues of any matrix is the same as the trace of a matrix, the proof is complete.\\
\clearpage
\section*{Appendix B: A toy simulation study}
\begin{figure}[h!]
    \centering
    \includegraphics[scale=0.52]{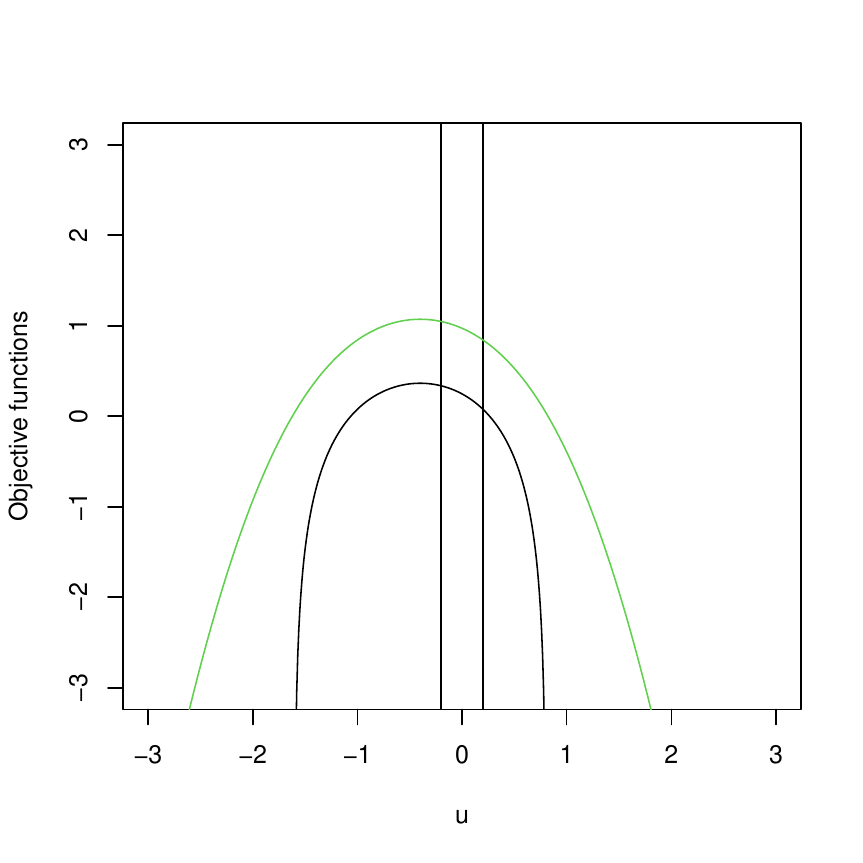}
    \caption{Objective functions for two considered optimization problems where the green one is related to problem \eqref{e00002}. $\alpha=0.4$ and $\lambda=0.5$ are considered, and two vertical lines show the acceptable area. }
    \label{05}
\end{figure}

For the given conjecture in \eqref{eq13}, we simulate the data from a multivariate Gaussian distribution $\mathcal{N}_p(\bm \mu,\bm \Sigma)$, where $\bm \mu$ is zero vector and $\bm \Sigma=[\sigma_{k,k^{'}}]$ with $\sigma_{k,k}=1$ and $\sigma_{k,k^{'}}=0.6^2$ for $k\neq k^{'}$. Then we compare the solution of two optimization problems \eqref{e00002} and $\max\log \det (\bm S+\bm U)$ subject to $||\bm U||_\infty\leq \lambda\alpha$ for some sample covariance matrices simulated from multivariate Gaussian distribution. It should be noted for the case $p=2$. They have the same solution. Two objective functions for $\alpha=0.4$ and $\lambda=0.6$ plot in Figure \ref{05}, and it is clear they have the same optimal point.\\
For $p=3$, if we consider $\alpha=0.4$ and $\lambda=0.6$ and matrix $U=[u_{k,k^{'}}]$, Table \ref{ta1} shows the difference between the estimated non-diagonal elements $u_{12}, u_{13}$ and $u_{23}$ of the solution of two optimization problems  for $15$ iterations. The differences show the results are very close to zero, as we expect in the proposed conjecture.
\begin{table}[h!]\label{ta1}
\begin{center}
    \caption{The difference between the non-diagonal elements of solution of two optimization problems}
    \vspace{0.6cm}
\begin{tabular}{|c|c|}
\hline
$Iteration$ & $p=3$ \\ \hline
1 & 4.191486e-11, 1.373684e-08, 5.860205e-11 \\ \hline
2 & 2.810245e-09, 1.942890e-16, 2.380667e-09 \\ \hline
3 & 1.134227e-06, 6.891155e-07, 2.775558e-17 \\ \hline
4 & 1.354186e-10, 1.921179e-08, 8.326673e-17 \\ \hline
5 & 1.897935e-07, 1.223164e-07, 0.000000e+00 \\ \hline
6 & 1.767697e-07, 3.756689e-08, 3.608225e-16 \\ \hline
7 & 2.021518e-07, 6.238265e-08, 4.440892e-16 \\ \hline
8 & 6.032975e-08, 2.808495e-08, 1.665335e-16 \\ \hline
9 & 1.342834e-06, 7.493853e-07, 5.551115e-17 \\ \hline
10 & 8.326673e-17, 7.227610e-10, 9.018657e-10 \\ \hline
11 & 5.024786e-08, 5.419546e-06, 3.608225e-16 \\ \hline
12 & 4.345127e-07, 1.283257e-07, 5.551115e-17 \\ \hline
13 & 3.330669e-16, 8.649000e-10, 2.368810e-10 \\ \hline
14 & 5.007422e-11, 2.035536e-09, 4.718448e-16 \\ \hline
15 & 4.617334e-12, 8.790247e-03, 1.942890e-16 \\ \hline
\end{tabular}
\end{center}
\end{table}
\clearpage
\end{document}